\documentclass[superscriptaddress,secnumarabic,amssymb,amsmath,nobibnotes,aps,prd,showkeys,nofootinbib,onecolumn,12pt]{revtex4}
\usepackage[utf8]{inputenc} 
\usepackage[T1]{fontenc}    
\usepackage{hyperref}       
\usepackage{url}            
\usepackage{booktabs}       
\usepackage{amsfonts}       
\usepackage{nicefrac}       
\usepackage{microtype}      
\usepackage{lipsum}		
\usepackage{graphicx}
\usepackage{natbib}
\usepackage{doi}
\usepackage{graphicx}
\usepackage{subfigure}
\usepackage{epsf}
\usepackage{bm}
\usepackage{amsmath}
\usepackage{amsfonts}
\usepackage{amssymb}
\usepackage{graphicx}
\usepackage{tabularx}
\usepackage{color}%
\providecommand{\U}[1]{\protect\rule{.1in}{.1in} }

\newcommand{\be}{\begin{equation}}
\newcommand{\ee}{\end{equation}}

\newcommand{\mincir}{\raise
-3.truept\hbox{\rlap{\hbox{$\sim$}}\raise4.truept\hbox{$<$}\ }}
\newcommand{\magcir}{\raise
-3.truept\hbox{\rlap{\hbox{$\sim$}}\raise4.truept\hbox{$>$}\ }}

\providecommand{\U}[1]{\protect\rule{.1in}{.1in}}

\usepackage{tikz,xcolor,hyperref}

\definecolor{lime}{HTML}{A6CE39}
\DeclareRobustCommand{\orcidicon}{%
	\begin{tikzpicture}
	\draw[lime, fill=lime] (0,0) 
	circle [radius=0.16] 
	node[white] {{\fontfamily{qag}\selectfont \tiny ID}};
	\draw[white, fill=white] (-0.0625,0.095) 
	circle [radius=0.007];
	\end{tikzpicture}
	\hspace{-2mm}
}

\foreach \x in {A, ..., Z}{%
	\expandafter\xdef\csname orcid\x\endcsname{\noexpand\href{https//orcid.org/\csname orcidauthor\x\endcsname}{\noexpand\orcidicon}}
}

\begin{document}

\title{Revisiting Fractional Cosmology}

\newcommand{\orcidauthorA}{0000-0002-9187-4402}
\newcommand{\orcidauthorB}{0000-0002-1625-0624}
\newcommand{\orcidauthorC}{0000-0002-1152-6548}
\newcommand{\orcidauthorD}{0000-0001-5852-514X}
\newcommand{\orcidauthorE}{0000-0002-9966-5517}

\author{Bayron Micolta-Riascos\orcidA{}}
\email{bayron.micolta@alumnos.ucn.cl}
\affiliation{Departmento de Física, Universidad Católica del Norte, Av. Angamos 0610,  
Antofagasta 1270709, Chile}

\author{Alfredo D.  Millano\orcidB{}}
\email{alfredo.millano@alumnos.ucn.cl}
\affiliation{Departamento de Matem\'{a}ticas, Universidad Cat\'{o}lica del Norte, Avda.
Angamos 0610, Casilla 1280, Antofagasta 1270709, Chile} 

\author{Genly Leon\orcidC{}}
\email{genly.leon@ucn.cl}
\affiliation{Departamento de Matem\'{a}ticas, Universidad Cat\'{o}lica del Norte, Avda.
Angamos 0610, Casilla 1280, Antofagasta 1270709, Chile}
\affiliation{Institute of Systems Science, Durban University of Technology, P.O. Box 1334,
\mbox{Durban 4000,  Republic of South Africa}}

\author{Cristián Erices\orcidD{}}
\email{cristian.erices@ucentral.cl}
\affiliation{Vicerrectoría Académica,  Universidad  Central de Chile, Toesca 1783, Santiago 8320000, Chile}
\affiliation{Departamento de Matemática, Física y Estadística, Universidad Católica del Maule, Av. San Miguel 3605, Talca 3480094, Chile}

\author{Andronikos Paliathanasis\orcidE{}}
\email{anpaliat@phys.uoa.gr}
\affiliation{Institute of Systems Science, Durban University of Technology, P.O. Box 1334,
\mbox{Durban 4000,  Republic of South Africa}}
\affiliation{Departamento de Matem\'{a}ticas, Universidad Cat\'{o}lica del Norte, Avda.
Angamos 0610, Casilla 1280, Antofagasta 1270709, Chile}

\begin{abstract}
Recently, the research community has been exploring fractional calculus to address problems related to cosmology; in this approach, the gravitational action integral is altered, leading to a modified Friedmann equation, then the resulting theory is compared against observational data. In this context, dynamical systems can be used along with an analysis the phase spaces for different values of the fractional order of the derivative and their different matter contents. The equilibrium points are classified, providing a range for the order of the fractional derivative in order to investigate whether the cosmological history can be reconstructed and a late-time accelerating power-law solution obtained for the scale factor. In this paper, we discuss the physical interpretation of the corresponding cosmological solutions with particular emphasis on the influence of the fractional order of the derivative in a theory of gravity that includes a scalar field minimally coupled to gravity. The presented results improve and extend those obtained previously, further demonstrating that fractional calculus can play a relevant role in cosmology.
\end{abstract}


\date{\today}
\maketitle

\section{Introduction}
In contemporary cosmology, the matter of the universe is made up of baryonic matter, photons, neutrinos, dark matter, and dark energy. In particular, in $\Lambda$CDM Cosmology, the dark energy component is a cosmological constant ($\Lambda$) and cold dark matter (CDM) is present. $\Lambda$CDM  describes the late-time acceleration of the universe observed from type Ia supernovae (SnIa) \cite{Riess:1998} and confirmed by the Cosmic Microwave Background Radiation (CMBR)~\cite{Planck:2018}. It describes the structural formation of the universe and has excellent agreement with observations. However, the model suffers from the well-known cosmological constant problem \cite{Zeldovich, Weinberg}, and the origin of the late-time acceleration of the universe remains to be discovered \cite {Carroll:2000}. More recently, this was coined the $H_0$-tension problem, which states that the value of the Hubble constant as measured by local SH0ES observations \cite{Riess:2019cxk} is in tension with the value estimated from the Planck \cite{Planck:2018} observations. A possible alternative that could resolve this tension is to consider extensions of $\Lambda$CDM \cite{DiValentino:2021izs}. Common approaches fall into two main categories: (i) assuming a dark energy fluid which affects the acceleration of the universe or (ii) modifying General Relativity to obtain cosmic acceleration without adding dark energy. 
Noncommutative theories, quantum cosmology, quantum deformation, deformed phase space,  Brans--Dicke theory, and noncommutative minisuperspace are among the alternatives to the cosmological constant that have been proposed; for detailed examples, see \cite{Rasouli:2013sda, Jalalzadeh:2014jea, Rasouli:2014dxa, Rasouli:2014dba, Rasouli:2016xdo, Rasouli:2016syh, Jalalzadeh:2017jdo} and references therein. Scalar field theories of particular interest include  \cite{Jordan:1958zz,Brans:1961sx,Horndeski:1974wa,Guth:1980zm,Ibanez:1995zs,Coley:1997nk,Coley:1999mj,Coley:2000zw,Coley:2000yc,Rubano:2001su,Coley:2003tf,Elizalde:2004mq,Guo:2004fq,Feng:2004ff,Capozziello:2005tf, Capozziello:2005mj,Urena-Lopez:2005pzi,Nojiri:2005pu,Zhang:2005eg,Zhang:2005kj,Briscese:2006xu,Nojiri:2006ww,Lazkoz:2006pa,Lazkoz:2007mx,Setare:2008pz,Setare:2008pc,Elizalde:2008yf,Basilakos:2011rx,Ito:2011ae,Frampton:2011rh,Leon:2012vt,Xu:2012jf,Leon:2012mt,Chervon:2013btx,Paliathanasis:2015gga,Barrow:2016qkh,Barrow:2016wiy,Paliathanasis:2017ocj,Tsamparlis:2018nyo,Mishra:2018dzq,Barrow:2018zav,Quiros:2019ktw,Marciu:2020vve,Dimakis:2020tzc,Paliathanasis:2020wjl,Banerjee:2020xcn,Lee:2022cyh,Motta:2021hvl,Astashenok:2012kb,Astashenok:2012tv,Bamba:2014daa,Odintsov:2018zai,Odintsov:2018uaw,Paliathanasis:2014yfa}.

This paper focuses on the second approach under the formalism known as fractional calculus. This consists of a generalization of classical integer order calculus to a form with derivatives and integrals of arbitrary (real or complex) order \cite{Tarasov2013}. This formalism has drawn increasing attention in the study of so-called ``anomalous" social and physical behaviours, in which the scaling power law of fractional order appears universal as an empirical description of such complex phenomena. 
The standard mathematical models of integer-order derivatives, including nonlinear models, need to be revised in many cases where the power law is observed. In order to accurately reflect the nonlocal frequency- and history-dependent properties of power law phenomena, alternative modelling tools such as fractional calculus have to be introduced. Research into fractional differentiation is inherently multi-disciplinary, has applications across various disciplines, and in general is an excellent research activity. Relevant texts on this topic include \cite{monje2010fractional,bandyopadhyay2014stabilization,padula2014advances,herrmann2014fractional,tarasov2019applications,klafter2012fractional,malinowska2015advanced,lorenzo2016fractional}. Specific areas of interest include fractional quantum mechanics and gravity for fractional spacetime \cite{Calcagni:2010bj, Calcagni:2009kc} and fractional quantum field theory at positive temperature \cite{Lim:2006hp, LimEab+2019+237+256}. Other applications of Quantum Cosmology can be found in \cite{VargasMoniz:2020hve, Moniz:2020emn, Rasouli:2021lgy, Jalalzadeh:2021gtq}. In addition, fractional calculus has recently been explored to address problems related to cosmology in \cite{Shchigolev:2010vh, Shchigolev:2012rp, Shchigolev:2013jq, Calcagni:2013yqa, Shchigolev:2015rei, Calcagni:2016azd, Shchigolev:2021lbm, Jalalzadeh:2022uhl, Calcagni:2020ads, Calcagni:2021ipd, Calcagni:2021aap, Calcagni:2020tvw, Calcagni:2019ngc, Calcagni:2017via, Calcagni:2016ofu, El-Nabulsi:2012wpc, El-Nabulsi:2015szp, Jamil:2011uj, El-Nabulsi:2013mma, El-Nabulsi:2013mwa, Rami:2015kha, El-Nabulsi:2016dsj, El-Nabulsi:2017vmp, El-Nabulsi:2017jss, Debnath2012, Debnath2013, Roberts:2009ix, Vacaru:2010fb, Vacaru:2010wj, Vacaru:2010wn, Garcia-Aspeitia:2022uxz}. 

Modified cosmological equations of fractional cosmology were tested against data from cosmic chronometers and observations of type Ia supernovae in \cite{Garcia-Aspeitia:2022uxz}. A joint analysis allowed the range to be narrowed to the fractional order of the derivative. Furthermore, a dynamical system was presented and a stability analysis was carried out by introducing dimensionless variables and solving the Friedmann constraint locally around the equilibrium points. Finally, a range of the fractional order of the derivative was arranged in order to obtain a late-term accelerating power-law solution for the scale factor. Finally, the physical interpretation of the corresponding cosmological solution was discussed.

The natural generalization of the model studied in \cite{Garcia-Aspeitia:2022uxz} is, investigating the influence of the fractional order of the derivative in a fractional theory of gravity, including a scalar field minimally coupled to gravity. Below, we review known results and discuss new results in the context of cosmologies with a scalar field used in the fractional formulation of gravity. 
According to our research, it is possible to obtain relevant information on the properties of the flow associated with autonomous systems of ordinary differential equations from the cosmological context through the use of qualitative techniques of the theory of dynamical systems. In particular, combining local and global variables allows cosmologies with a scalar field to be qualitatively described in the context of fractional calculus. In addition, it is possible to provide precise schemes for finding analytical approximations of the solutions and exact solutions by choosing various approaches. Finally, we consider corrections of the Friedmann equation based on fractional calculus formalism, which describes inflationary cosmologies with a scalar field using the Friedmann--Lema\^{i}tre--Robertson--Walker and Bianchi I metrics. Bianchi I spacetime is the simplest homogeneous and anisotropic model. The limit of isotropization is reduced to the FLRW metric. Another essential characteristic of the Bianchi I Universe is that the Kasner Universe is recovered in the case of the vacuum in GR. The latter describes the evolution of the Mixmaster Universe near the cosmological singularity. While our universe is isotropic, anisotropies played an important role in its early history; hence, studying the evolution of anisotropies in fractional calculus is particularly interesting.

The primary approach uses dynamical systems to determine states and asymptotic solutions \cite{wainwrightellis1997}. This study consists of several steps: determining equilibrium points, linearization in their neighbourhood, finding the eigenvalues of the associated Jacobian matrix, checking the stability conditions in the neighbourhood of the equilibrium points, finding the sets of stability and instability and determining the basin of attraction, etc. Lyapunov's stability theorem is the most general result for determining the asymptotic stability of an equilibrium point. As far as we know, few works have used the Lyapunov method in cosmology \cite{Setare:2010zd, Cardoso:2008bp, Lavkin:1990gu, Charters:2001hi, Arefeva:2009tkq}. The Lyapunov stability method requires the use of the strict Lyapunov function, the construction of which is laborious, though not impossible. The Hartman--Grobman theorem (Theorem 19.12.6 in \cite{wiggins2006introduction} p. 350) can be used to investigate the stability of hyperbolic equilibrium points of nonlinear autonomous vector fields from the linearized system near the equilibrium point. For isolated non-hyperbolic equilibrium points, the normal forms theorem (Theorem 2.3.1 in \cite{arrowsmith1990introduction}) can be used, which contains the Hartman--Grobman theorem as a particular case. The normal forms of the dynamical system can have periodic solutions for a  broad set of initial conditions, implying that an initially expanding closed isotropic universe can exhibit oscillatory behaviour \cite{Leon:2009rc, Miritzis:2007yn}. On the other hand, the invariant manifold theorem (Theorem 3.2.1 in \cite{wiggins2006introduction}) affirms the existence of stable and unstable local manifolds under suitable conditions for the vector field. However, it only allows  partial information about the stability of equilibrium points to be obtained, and does not provide a method for determining the stability or instability of manifolds. 

For investigation of the asymptotic states of the system, the appropriate concepts are the $\alpha$ and $\omega$- limit sets of $x\in \mathbb{R}^n$, that is, the past and future attractors of $x$, respectively (see Definition 8.1.2 in \cite{ wiggins2006introduction} p.105). To characterize these invariant sets, the LaSalle Invariance Principle (\cite{LASALLE196857}; Theorem 8.3.1 \cite{wiggins2006introduction}, p. 111) or Monotonicity Principle (\cite{wainwrightellis1997}, p. 103; \cite{LeBlanc:1994qm} p. 536) can be used. When applying the Monotonicity Principle a monotonic function is required; in certain cases, this is suggested by the Hamiltonian formulation of the field equations~\cite{Heinzle:2009zb}. Furthermore, the Poincar\'e-Bendixson \cite{Coley:2003mj} theorem can be used in $\mathbb{R}^2$. Its corollary can distinguish between all of the possible $\omega$-limit sets of the plane. Then, any compact asymptotic set is one of the following: (1) an equilibrium point, (2) a periodic orbit, or (3) the union of equilibrium points and heteroclinic or homoclinic orbits. If a closed orbit (i.e., periodic, heteroclinic, or homoclinic) can be ruled out, all asymptotic behaviour corresponds to an equilibrium point. For this purpose, Dulac's criteria can be used (Theorem 3 \cite{Coley:1999uh} p. 6, \cite{wainwrightellis1997}, p. 94, and \cite{Coley:2003mj}) based on the construction of a Dulac function. Dynamical systems tools and observational tests have been explored and applied in various cosmological contexts \cite{Hernandez-Almada:2020uyr, Leon:2021wyx, Hernandez-Almada:2021rjs, Hernandez-Almada:2021aiw, Garcia-Aspeitia:2022uxz}. These methods have proven to be a robust scheme for investigating the physical behaviour of cosmological models, and can be used in new contexts such as in this paper.

 There are currently several definitions of the fractional derivative, including the Riemann-Liouville and Caputo derivatives, among others \cite{Uch:2013}. 
The Caputo left derivative is defined by
\begin{align}
 & { }^{C} D_{t}^{\mu} f(t)  = \frac{1}{\Gamma(n-\mu)} \int_{c}^{t} \frac{\frac{d^{n}}{d \theta^{n}} f(\theta)}{(t-\theta)^{\mu-n+1}} d \theta, \; \text{where}\; n=\left\{\begin{array}{cc}
     [\mu] +1& \mu \notin \mathbb{N}\\
     \mu & \mu \in \mathbb{N}
\end{array} \right.,
\end{align}
where $\Gamma(\cdot)$ is the Gamma function.

The following relation for second-order derivatives generalizes the rule of successive derivatives \cite{Uch:2013}:
\begin{equation}
D_{t}^{\mu} \left[D_{t}^{\beta} f(t)\right]=D_{t}^{\mu+\beta} f(t)-\sum_{j= 1}^{n} D_{t}^{\beta-j} f(c+) \frac{(t-c)^{-\mu-j}}{\Gamma(1-\mu-j)}.
\end{equation}

Additionally, Leibniz's rule \cite{Uch:2013} is written as
\begin{equation}
D_{t}^{\mu}[f(t) g(t)]=\sum_{k=0}^{\infty} \frac{\Gamma(\mu+1)}{k ! \Gamma(\mu-k+1)} D_{t}^{\mu-k} f(t) D_{t}^{k} g(t),
\end{equation}
recovering the usual rule when $\mu=n\in \mathbb{N}$.

This remainder of this work is organized as follows. 
 An analytical solution to the fractional Friedmann equation is discussed in Section \ref{FracFriedmann}. In Section \ref{alternattive_I}, an alternative study is presented that uses Riccati's Equation \eqref{NewIntegrableH}, assuming that the matter components have the equation of state $p_i=w_i\rho_i$, where $w_i \neq -1$ are constants. In Section \ref{Bianchi}, the Bianchi I Cosmology is examined in phase space. In Section \ref{AlternattiveII}, an alternative study is carried out for the Bianchi I metric using the Riccati Equation \eqref{NewIntegrableH}, where it is assumed that the equation of state of the matter components is $ p_i= w_i\rho_i$, with $w_i \neq -1$ being constants. Section \ref{ch_4} presents the fractional formulation of a cosmology with a scalar field and an additional matter source. Here, we generalize the results from Section \ref{sec:cosmology}. Section \ref{RRResults} summarizes the most relevant results, and our conclusions are presented in Section \ref{conclusiones}.

\section{Cosmological Model in Fractional Formulation}

The variational approach with fractional action was developed by, e.g., \cite{El-Nabulsi:2005oyu, El-Nabulsi:2007lla, El-Nabulsi:2007wgc, El-Nabulsi:2008oqn, Roberts:2009ix,10.5555/1466940.1466942}. 
With the following fractional action integral:
\begin{align}
 S
   & = \frac{1}{\Gamma(\mu)}\int_0^t \mathcal{L}\left(\theta, q_i(\theta), \dot{q}_i(\theta), \ddot{q}_i(\theta)\right)(t-\theta)^{\mu-1} d\theta, \label{GenCFL}
\end{align}
\noindent 
where $\Gamma(\mu)$ is the Gamma function, $\mathcal{L}$ is the Lagrangian, $\mu$ is the constant fractional parameter, and $t$ and $ \theta$ are the physical and intrinsic time, respectively, variation of \eqref{GenCFL} with respect to $q_i$ leads to the Euler--Poisson equations \cite{10.5555/1466940.1466942}:  

  \begin{align}
&   \frac{\partial \mathcal{L}\left(\theta, q_i(\theta), \dot{q}_i(\theta), \ddot{q}_i(\theta)\right)}{\partial q_i}   -\frac{d}{d \theta} \frac{\partial \mathcal{L}\left(\theta, q_i(\theta), \dot{q}_i(\theta), \ddot{q}_i(\theta)\right) }{\partial \dot{q}_i}  + \frac{d^2}{d \theta^2} \frac{\partial \mathcal{L}\left(\theta, q_i(\theta), \dot{q}_i(\theta), \ddot{q}_i(\theta)\right) }{\partial \ddot{q}_i}  \nonumber \\
& =  \frac{1-\mu}{t-\theta} \left(\frac{\partial \mathcal{L}\left(\theta, q_i(\theta), \dot{q}_i(\theta), \ddot{q}_i(\theta)\right)}{\partial \dot{q}_i}   -2 \frac{d}{d \theta} \frac{\partial  \mathcal{L}\left(\theta, q_i(\theta), \dot{q}_i(\theta), \ddot{q}_i(\theta)\right)}{\partial \ddot{q}_i} \right)  \nonumber \\
& - \frac{\left(1-\mu\right)\left(2-\mu\right)}{\left(t-\theta\right)^2} \frac{\partial  \mathcal{L}\left(\theta, q_i(\theta), \dot{q}_i(\theta), \ddot{q}_i(\theta)\right)}{\partial \ddot{q}_i}. \label{EP}
 \end{align}

\subsection{Flat FLRW Fractional Model}\label{sec:cosmology}
In cosmology, it is assumed that the geometry of spacetime is provided by the flat Friedmann--Lema\^{i}tre--Robertson--Walker (FLRW) metric: 
\begin{equation}
  ds^2=-N^2 (t) dt^2+ a^2(t)(dx^2+ dy^2+dz^2),  
\end{equation}
where  $a(t)$ denotes the scale factor.
This result is based on Planck's observations \cite{Planck:2018}.
The effective fractional action used in \cite{Garcia-Aspeitia:2022uxz} is
\begin{align}
 S_{\text{eff}}
   & = \frac{1}{\Gamma(\mu)}\int_0^t \Bigg[\frac{3}{8\pi G}\Bigg(\frac{a^2(\theta ) \ddot{a}(\theta )}{N^2(\theta )} +\frac{a(\theta ) {\dot{a}}^2(\theta )}{N^2(\theta )}-\frac{a^2(\theta ) \dot{a}(\theta ) \dot{N}(\theta )}{N^3(\theta)}\Bigg)  +a^3(\theta)\mathcal{L}_{\text{m}}\Bigg](t-\theta)^{\mu-1} N(\theta) d\theta, \label{CFL}
\end{align}
\noindent
where $\Gamma(\mu)$ is the Gamma function, $\mathcal{L}_{\text{m}}$ is the matter Lagrangian, $\mu$ is the constant fractional parameter, and $t$ and $ \theta$ are the physical and intrinsic time, respectively \cite{Shchigolev:2010vh}. The Euler--Poisson Equations \eqref{EP} obtained after varying the action \eqref{CFL} for $q_i\in \{N, a \}$ lead to the field equations 
\begin{align}
&\left(\frac{\dot{a}(\theta )}{a(\theta )}\right)^2 + \frac{(1-\mu)}{(t-\theta )} \frac{\dot{a}(\theta )}{a(\theta )}= \frac{8 \pi G}{3} \rho(\theta), \label{FEQ1}\\
&    \frac{\ddot{a}(\theta )}{a(\theta )}+\frac{1}{2}\left(\frac{\dot{a}(\theta )}{a(\theta )}\right)^2+ \frac{(1-\mu)}{(t-\theta )} \frac{\dot{a}(\theta )}{a(\theta )}+\frac{(\mu -2) (\mu -1)}{2 (t-\theta )^2}=- 4 \pi G p (\theta). \label{REQ1}
\end{align}
\noindent
where $\rho(\theta)$ and $p(\theta)$ are the total energy density and the isotropic pressure of the matter fields; here, we have substituted the lapse function $N=1$ after the variation. To designate the temporary independent variables, the rule $t-\theta \mapsto t$, $\theta \mapsto t$ \cite{Shchigolev:2010vh} is used, where the dots denote these derivatives. Furthermore, the Hubble parameter is defined as $H\equiv\dot{a}/a$.

Including all matter sources in Equations \eqref{FEQ1} and \eqref{REQ1}, after performing algebra the following Raychaudhuri equation (with $N(t)=1$) is obtained:
\begin{equation}
   \dot{H}+\frac{(\mu -1) H}{2 t}+\frac{(\mu -2) (\mu -1)}{2 t^2}=- 4 \pi G \sum_i (p_i+\rho_i), \label{IntegrableH}
\end{equation}
along with the Friedmann equation
\begin{equation}
    H^2+\frac{(1-\mu)}{t}H=\frac{8\pi G}{3}\sum_i\rho_i. \label{FriedmannFrac}
\end{equation}

Furthermore, the continuity equation leads to 
\begin{equation}
    \sum_i\left[\dot{\rho}_i+3\left(H+\frac{1-\mu}{3t}\right)(\rho_i+p_i)\right]=0,\label{CE}
\end{equation}
where $\rho_i$ and $p_i$ are the density and pressure of the $i$th matter component and the sum is over all species, e.g., matter, radiation, etc. Note that when $\mu=1$ in Formula \eqref{FriedmannFrac} and Formula \eqref{CE}, the standard cosmology without $\Lambda$ is recovered, which by itself does not produce an accelerated expanding universe.

Using the equation of state $p_i=w_i\rho_i$, where $w_i \neq -1$ are constants, we have
\begin{align}
       & \sum_i (1+w_i)\rho_i \left[\frac{\dot{\rho}_i}{ (1+w_i) \rho_i} + 3\frac{\dot{a}}{a}+ \frac {1-\mu}{t} \right] = \sum_i (1+w_i)\rho_i \frac{\mathrm{d}}{\mathrm{d}t}\left[ \ln \left( {\rho_i}^{ 1/(1+w_i)} a^3 t^{1-\mu}\right)\right].
\end{align}

Assuming separate conservation equations for each species and integrating for each ${\rho_i}$, we have the following solution:
\begin{align} 
\rho_{i}(t)= \rho_{0i} a(t)^{-3 (1+w_i)} \left(t/t_U\right)^{(\mu-1)(1+w_i) }. \label{EQ16}
\end{align}
where $a(t_U)=1$,  $t_U$ is the age of the universe and $\rho_{0i}$ is the current value of the energy density of the $i$th species. 
Therefore, by substituting \eqref{EQ16} into \eqref{FriedmannFrac}, we have
\begin{align}
   & H^2+\frac{(1-\mu)}{t}H =\frac{8\pi G}{3}\sum_i \rho_{0i} a^{-3 (1+w_i)}\left(t/t_U\right)^{(\mu-1)(1+w_i)}. \label{EH}
\end{align}

\subsection{Analytic Solution for the Fractional Friedmann Equation} \label{FracFriedmann}

Note that for $\mu\neq 1$, the modified continuity Equation \eqref{CE} provides the condition
\begin{align}
    \frac{8 \pi G}{3} \sum_i p_i=\frac{2 (\mu -3) H}{t}+H^2-\frac{(\mu -2) (\mu -1)} {t^2}. \label{Newpressure}
\end{align}

Combining these results with \eqref{IntegrableH}
and \eqref{FriedmannFrac}, we have the Riccati equation
\begin{equation}
   \dot{H}+\frac{2 (\mu -4) H }{t}+3 H ^2-\frac{(\mu -2) (\mu -1)}{t^2}=0. \label{NewIntegrableH}
\end{equation}

The analytical solution of \eqref{NewIntegrableH} (see an analogous case in \cite{Shchigolev:2012rp} Equation (36)) is as follows:
\begin{equation}
 H(t)=  \frac{9-2 \mu + r }{6 t} -\frac{r c_1}{3t(t^{r}+c_1)}, \label{solution}
\end{equation}
where
\begin{equation} \label{eq:c1_newH}
    c_1=\frac{t_U^{r} \left(-6 H_0 t_U-2 \mu +r+9\right)}{6 H_0 t_U+2 \mu +r-9}, \; r=\sqrt{8 \mu(2 \mu -9)+105}.
\end{equation}
Here, $c_1$ is an integration constant depending on $\mu$, the value $H_0$, and the age of the universe $t_U$.

The relation between the redshift $z$ and cosmic time $t$ is through the scale factor,
\begin{align}
a(z) & :=(1+z)^{-1}  = \left[\frac{t^{r}+c_1}{{t_U^{r}+c_1}}\right]^{\frac{1}{3}} \left[ \frac{t}{ t_U}\right]^{\frac{1}{6} \left(-2 \mu -r+9\right)}. \label{tz}
\end{align}

Then, for large $t$, the asymptotic scale factor can be expressed as
\begin{equation}
    a(t)\simeq t^{\frac{1}{6} \left(-2 \mu +r+9\right)}. \label{H(t)}
\end{equation}

Therefore, for large $t$ we need to have $q<0$, in which case we have late-time acceleration without adding dark energy.

Alternative expressions for $E$ and the deceleration parameter that make use of the exact solution of \eqref{NewIntegrableH} provided by \eqref{solution} include
\begin{align}
    E(t) & = \frac{1}{H_0} \left[\frac{9-2 \mu + r }{6 t} -\frac{r c_1}{3t(t^{r}+c_1)}\right], \label{alternativeE}
\end{align}
and 
\begin{align}
  q(t)= -1 & +\frac{-6 c_1{}^2 \left(2 \mu +r-9\right)}{\left(\left(-2 \mu +r+9\right ) t^{r} + c_1 \left(2 \mu +r-9\right)\right){}^2} \nonumber\\
  &+\frac{6 \left(-2 \mu +r+9\right) t^{2 r}}{\left(\left(-2 \mu +r+9\right) t^{r} + c_1 \left(2 \mu +r-9\right)\right){}^2} \nonumber \\
   &+\frac{-24 c_1 (\mu (8 \mu -35)+48) t^{r}}{\left(\left(-2 \mu +r+9\right) t^{r} + c_1 \left(2 \mu +r-9\right)\right){}^2}\label{Alternative-q}
\end{align}
where $c_1$ is defined by \eqref{H(t)}, the relation between $t$ and $z$ is obtained by inverting \eqref{tz}, and $ r=\sqrt{8 \mu (2 \mu -9)+105}$.

\subsection{Dynamical Systems and Stability Analysis}
\label{alternattive_I}
In reference \cite{Garcia-Aspeitia:2022uxz}, a coupled system $d\mathbf{X}/d\tau = \mathbf{F}(\mathbf{X})$ subject to a constraint $G(\mathbf{ X}) = \mathbf{0}$ ($\mathbf{X}$ being the reduced phase space variables) was studied. The equilibrium points, determined by the equations $\mathbf{F}(\mathbf{X}) = \mathbf{0}, \mathbf{G} (\mathbf{X} ) = \mathbf{0}$, are of central importance for this investigation. Calculating the gradient $\nabla\mathbf{G}(\mathbf{X})$, if $\nabla\mathbf{G}(\mathbf{X})|_P \neq 0$ then the constraint $\mathbf{G} (\mathbf{X}) = \mathbf{0}$ is solved locally, obtaining a lower-dimensional system following \cite{Hewitt:1992sk, Nilsson:1995ah, Goliath:1998mx}. 

Instead of continuing the discussion in reference \cite{Garcia-Aspeitia:2022uxz}, an alternative study is presented here that uses the Riccati equation \eqref{NewIntegrableH}, assuming that the matter components have the equation of state $ p_i=w_i\rho_i$, where $w_i \neq -1$ are constants.

It can be observed that Equations \eqref{FriedmannFrac} and \eqref{Newpressure} impose restrictions on the type of matter components in the universe, say, 
\begin{align}
    H^2+\frac{(1-\mu)}{t}H &=\frac{8\pi G}{3}\sum_i\rho_i,
\\
  \frac{2 (\mu -3) H}{t}+H^2-\frac{ (\mu -2) (\mu -1)}{t^2} &= \frac{8 \pi G}{3} \sum_i w_i \rho_i.
\end{align}

The second condition is obtained by imposing separated conservation equations for each matter component in the case $\mu\neq 1$. In the rest of this paper, we assume $\mu \notin\{1,2\}$.  

Then, defining the dimensionless variables
\begin{equation}
    \Omega_i= \frac{8 \pi G \rho_i}{3 H^2}, \quad A=tH, \label{AlternativevarsFLRW}
\end{equation}
we have the constraints 
\begin{align}
 1+\frac{(1-\mu)}{A} &=\sum_i\Omega_i, \label{lig1}
\\
1+ \frac{2 (\mu -3)}{A}-\frac{ (\mu -2) (\mu -1)}{A^2}& = \sum_i w_i \Omega_i.
\end{align}

With the new derivative $f^{\prime}=\dot{f}/H$, we obtain the following for $\mu \neq 1$:
\begin{align}
\Omega_i^{\prime}& =\Omega_i \left[
   (2 q-3 w_i-1) + (w_i+1) \left(1- \sum_i\Omega_i\right)\right] \label{EvolOmegas},
\\
 A^{\prime} & = 1- A (1+q),  \label{Evol-Age}
\end{align} 
where the deceleration parameter can be obtained from Equation \eqref{NewIntegrableH}:
\begin{equation}
q:= -1 - \frac{\dot{H}}{H^2}= 2 + \frac{2 (\mu -4) }{A}-\frac{(\mu -2) (\mu -1)}{A^2}.\label{decc}
\end{equation}

Here, restriction \eqref{lig1} is used to remove the ${(1-\mu)}A^{-1}:= \left(1- \sum_i\Omega_i\right)$ term originally appearing in \eqref{EvolOmegas}.

For comparison with the Standard Model, it is assumed that the components of the universe are CDM ($\rho_{1}=\rho_{\text{m}}, w_1=w_ {\text{m} }=0$) and radiation ($\rho_{2}=\rho_{\text{r}}, w_2=w_{\text{r}}=1/3$); furthermore, we have the constraints
\begin{align}
 1+\frac{(1-\mu)}{A} &=\Omega_{\text{m}} +\Omega_{\text{r}}, \label{LigA}
\\
1+ \frac{2 (\mu -3)}{A}-\frac{ (\mu -2) (\mu -1)}{A^2}& = \frac{1}{3} \Omega_ {\text{r}}. \label{LigB}
\end{align}

Dimensionless energy densities evolve according to
\begin{align}
\Omega_{\text{m}}^{\prime}& =\Omega_{\text{m}} (2 q-\Omega_{\text{m}}-\Omega_{\text{r}}), \\
\Omega_{\text{r}}^{\prime}& =\frac{2}{3} \Omega_{\text{r}} (3 q-2 \Omega_{\text{m}}-2 \Omega_{\text{r}}-1),
\end{align}
and the age parameter evolves according to \eqref{Evol-Age}, where $q$ is provided by \eqref{decc}. 

Scaling the time variable by the factor $A^2$, that is,
\begin{equation}
    \frac{d f}{d \tau}= A^2 \frac{d f}{d \ln a}, \label{NEW-TIME}
\end{equation}
the following dynamical system is obtained:
\begin{align}
\frac{d \Omega_{\text{m}}}{d \tau}& = \Omega_{\text{m}} \left(5 A \mu +A (3 A-17)-2 \mu ^ 2+6 \mu -4\right), \label{EQ3.45}\\
\frac{d \Omega_{\text{r}}}{d \tau}& =\frac{2}{3} \Omega_{\text{r}} \left(8 A
   \mu +A (3 A-26)-3 \mu ^2+9 \mu -6\right), \label{EQ3.46}\\
\frac{dA}{d \tau} & =A \left(-2 A \mu -3 (A-3) A+\mu ^2-3 \mu +2\right), \label{EQ3.47}
\end{align}
with the relation $\left(1- \sum_i\Omega_i\right)=(1-\mu)A^{-1}$ used to obtain decoupled equations.

The expression \eqref{LigA} is trivially a first integral of the system \eqref{EQ3.45}, \eqref{EQ3.46}, \eqref{EQ3.47}. However, expression \eqref{LigB} is a first integral of the system \eqref{EQ3.45}, \eqref{EQ3.46}, \eqref{EQ3.47} only if
\begin{equation}
    A \left(3 A^3+A^2 (5 \mu -17)+A (\mu (13 \mu -64)+81)-(\mu -2) (\mu -1) (5 \mu -8)\right)=0. \label{LigNewB}
\end{equation}

Table \ref{tab:P} shows the equilibrium points/sets of the system \eqref{EQ3.45}, \eqref{EQ3.46}, \eqref{EQ3.47} that satisfy the compatibility conditions \eqref{LigB} and \eqref{LigNewB}. Based on physical considerations, we do not examine the points with $A=0$, corresponding to $t H \rightarrow 0$. 

\begin{table}[ht!]

 \caption{Equilibrium points/sets of the system \eqref{EQ3.45}, \eqref{EQ3.46}, \eqref{EQ3.47} that satisfy the compatibility conditions \eqref{LigB} and \eqref{LigNewB}; NH means nonhyperbolic.}
    \label{tab:P}
    \setlength{\tabcolsep}{2.5mm}
    \begin{tabular}{cccccccccc}
    \toprule 
\textbf{Label}   &  \boldmath{$\Omega_{\text{m}}$} & \boldmath{$\Omega_{\text{r}}$} & \boldmath{$A$}  & \boldmath{$\mu$} & \boldmath{$q$}  & \boldmath{$\lambda_1$} &   \boldmath{$\lambda_2$} & \boldmath{$\lambda_3$} & \textbf{Stability} \\ \midrule 
$P_1$& $ 0 $&$ 0$ & $\frac{3}{2}$ & $\frac{5}{2}$ &$ -\frac{1}{3} $& $-\frac{15}{2}$ & $-3$ & $-\frac{3}{2}$ & Sink \\\midrule
$P_2$& $ -\frac{2}{3}$ & $0$ & $\frac{3}{2}$ & $\frac{7}{2}$ & $-\frac{1}{3}$ & $-\frac{21}{2}$ & $-1$ & $0$ & NH. 2D stable manifold \\\midrule
$P_3$& $ 0$ & $-\frac{6}{5}$ & $\frac{15}{8} $& $\frac{41}{8}$ &$-\frac{7}{15}$ &$-\frac{375}{16}$ &$\frac{15}{16}$ & $0 $  & Saddle \\\toprule
    \end{tabular}
   
\end{table}

The results are listed below:
\begin{enumerate}

\item $P_1:$ exists for $\mu=\frac{5}{2}$. The deceleration parameter is $q= -\frac{1}{3} $. Therefore, it represents an accelerating power-law cosmological solution. The equilibrium point is a sink.
\item $P_2:$ exists for $\mu=\frac{7}{2}$. The deceleration parameter is $q=-\frac{1}{3}$. Therefore, it represents an accelerating power-law cosmological solution. The equilibrium point is nonhyperbolic with a two-dimensional stable manifold.

\item $P_3:$ exists for $\mu=\frac{41}{8}$. The deceleration parameter is $q=-\frac{7}{15}$. Therefore, it represents an accelerating power-law cosmological solution. The equilibrium point is a saddle.

\end{enumerate}

For equilibrium points with constant $A$, the corresponding cosmological solution is a power-law solution with scale factor $a(t)= \left(t/t_U\right)^{A}$. Then, the solutions $P_1$ and $P_2$ verify that $a(t)= \left(t/t_U\right)^{\frac{3}{2}}$. Finally, the solution $P_3$ satisfies $a(t)= \left(t/t_U\right)^\frac{15}{8}$.
Points $P_2$ and $P_3$ are nonphysical, as they lead to $\Omega_{\text{m}}<0$ and $\Omega_{\text{r}}<0$, respectively. 

Therefore, from the compatibility conditions of the problem and the condition $\mu \notin\{1,2\}$, and considering that the matter sources are radiation and cold dark matter, conditions are imposed on the parameter $\mu$, which can take the discrete values $\{5/2, 7/2, 41/8\}$ at the equilibrium points. The system is then reduced to a one-dimensional system provided by \eqref{EQ3.47} for $\mu\in \{5/2, 7/2, 41/8\}$.

\begin{figure}[ht!]
  \includegraphics[width=0.3\textwidth]{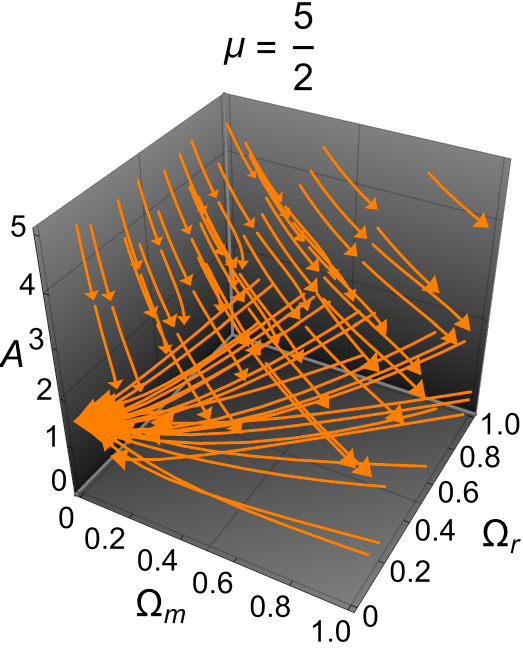}
  \includegraphics[width=0.3\textwidth]{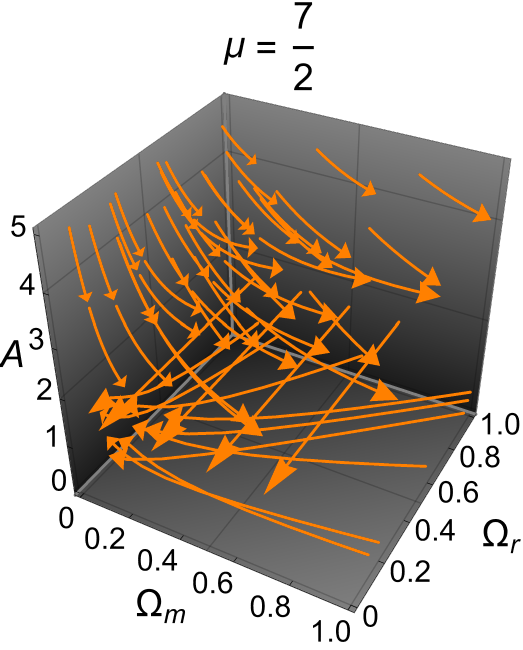}
  \includegraphics[width=0.3\textwidth]{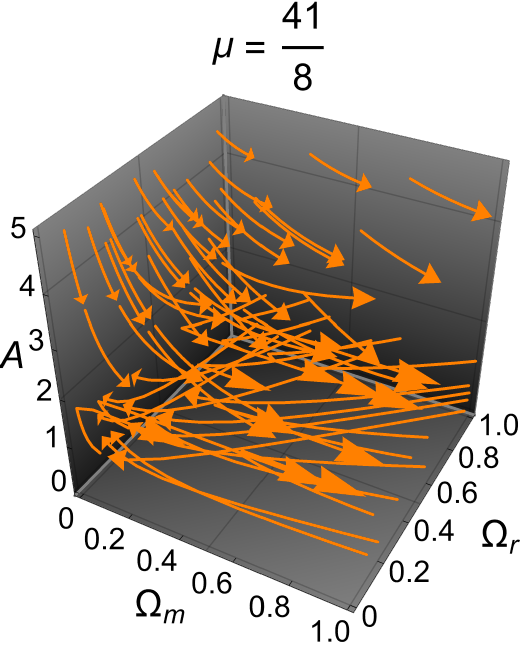}
 \caption{Flow of the system \eqref{EQ3.45}, \eqref{EQ3.46} and \eqref{EQ3.47} for the values of $\mu= 5/2, 7/2, 41/8$. }
 \label{fig:Flujos FLRW distintos mu}
\end{figure}

Figure \ref{fig:Flujos FLRW distintos mu} shows the system's flow for values  $\mu=\frac{5}{2}, \frac{7}{2}, \frac{41}{8}$.
\unskip

\begin{table}[ht!]
 \caption{Equilibrium points of the one-dimensional system \eqref{EQ3.47}, where $r=\sqrt{8\mu(2\mu-9)+105}$.}
    \label{tab:my_labelFLRW}
  \setlength{\tabcolsep}{8.5mm}
     \begin{tabular}{ccccc}
    \toprule
\textbf{Labels} & \boldmath{$A$} & \boldmath{$q$} & \textbf{Stability} \\\midrule
$Q_1$ & 0   & \text{Indeterminate} & Source \\\midrule
$Q_2$ &  $\frac{1}{6} \left(-r-2 \mu +9\right)$ & $ -\frac{2 (\mu -4) \mu +r+13}{2 (\mu -2) (\mu -1)}$&  Sink \\\midrule
$Q_3$ & $ \frac{1}{6} \left(r-2 \mu +9\right)$ & $ \frac{-2 (\mu -4) \mu +r-13}{2 (\mu -2) (\mu -1)}$ & Sink \\\bottomrule
 \end{tabular}   
\end{table}
Table \ref{tab:my_labelFLRW} shows the equilibrium points of the one-dimensional system \eqref{EQ3.47}, where $r=\sqrt{8\mu(2\mu-9)+105}$.


\begin{figure}[ht!]
    \includegraphics[scale=0.4]{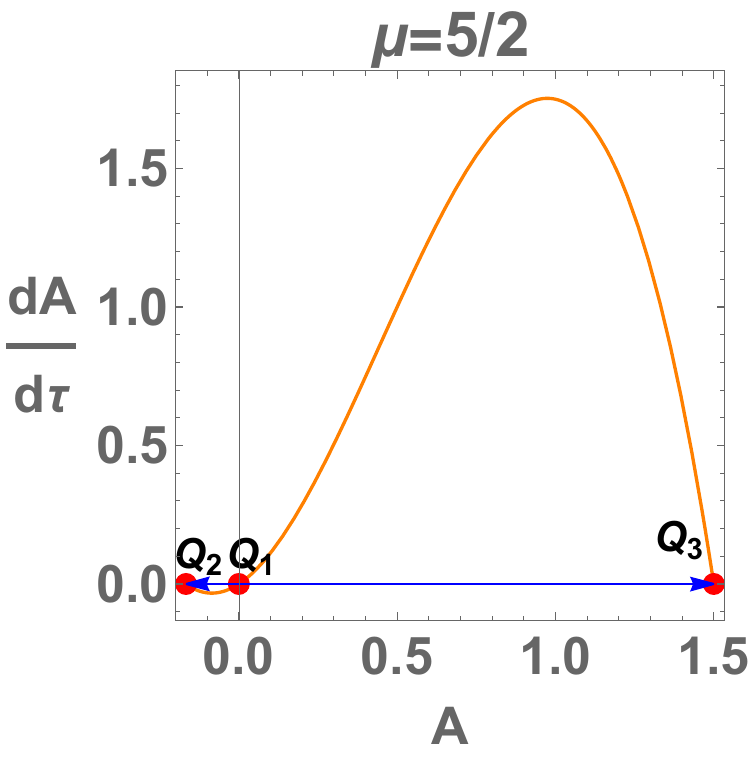}
    \includegraphics[scale=0.4]{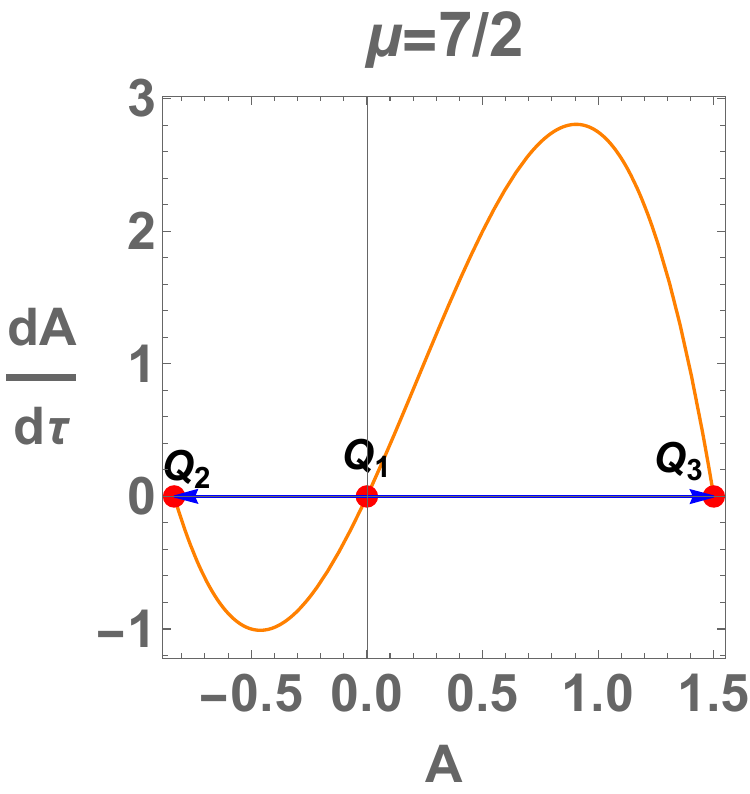}
    \includegraphics[scale=0.4]{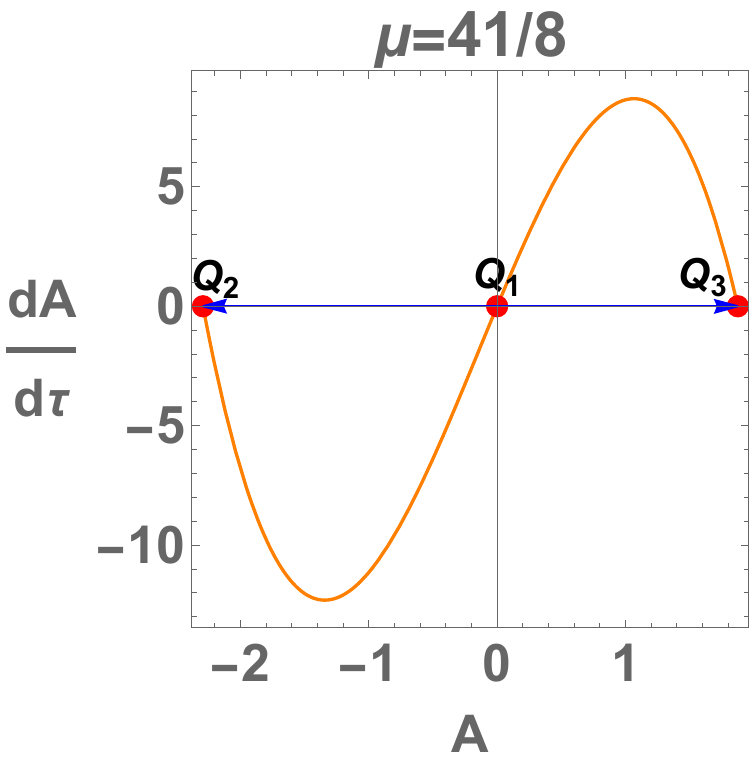}
    \caption{Flow of the one-dimensional system \eqref{EQ3.47} for $\mu=\frac{5}{2}, \frac{7}{2}, \frac{41}{8}$. Note that $Q_2$ is not a physical point, and the late-time attractors are $Q_2$ and $Q_3$.}
    \label{fig:1D-Alf}
\end{figure}
Figure \ref{fig:1D-Alf} displays the flow of the one-dimensional system \eqref{EQ3.47} for $\mu=\frac{5}{2}, \frac{7}{2}, \frac{41}{8}$. Note that $Q_2$ is not a physical point,  and the late-time attractors are $Q_2$ and $Q_3$.

A second alternative formulation is the following: the equation of state of one of the matter sources is not imposed; instead, it is deduced from the compatibility conditions. That is, it is assumed that the components of the universe are CDM and a fluid with a constant state equation to be determined ($\rho_{2}=\rho_{\text{X}}, w_2=w_{\text{X}}$).
Then, we have the constraints 
\begin{align}
 1+\frac{(1-\mu)}{A} &=\Omega_{\text{m}} +\Omega_{\text{X}}, \label{Lig2A}
\\
1+ \frac{2 (\mu -3)}{A}-\frac{ (\mu -2) (\mu -1)}{A^2}& = w_{\text{X}} \Omega_ {\text{X}}. \label{Lig2B}
\end{align}
The dimensionless energy densities are obtained according to
\begin{align}
\Omega_{\text{m}}^{\prime}& =\Omega_{\text{m}} (2 q-\Omega_{\text{m}}-\Omega_{\text{X}}), \\
\Omega_{\text{X}}^{\prime}& =\Omega_{\text{X}} \left[ (2 q-3 w_{\text{X}} -1) + (w_{\text {X}} +1) \left(1 -\Omega_{\text{m}}-\Omega_{\text{X}}\right)\right],
\end{align}
and the age parameter evolves according to \eqref{Evol-Age}, where $q$ is provided by \eqref{decc}. 

Using derivative \eqref{NEW-TIME}, the following dynamical system is obtained:
\begin{align}
\frac{d \Omega_{\text{m}}}{d \tau}& = \Omega_{\text{m}} \left(5 A \mu +A (3 A-17)-2 \mu ^ 2+6 \mu -4\right), \label{EQ3.61}\\
\frac{d \Omega_{\text{X}}}{d \tau}& = \Omega_{\text{X}} \left(A \mu (w_{\text{X}}+5)-A (3 A (w_{\text{X}}-1)+w_{\text{X}}+17)-2 \mu ^2+6 \mu -4\right), \label{EQ3.62} \\
\frac{d A}{d \tau}& = A \left(-2 A \mu -3 (A-3) A+\mu ^2-3 \mu +2\right), \label{EQ3.63}
\end{align}
with the relation $\left(1- \sum_i\Omega_i\right)=(1-\mu)A^{-1}$ used to obtain decoupled equations.

Expression \eqref{Lig2A} is trivially a first integral of the system \eqref{EQ3.61}, \eqref{EQ3.62}, \eqref{EQ3.63}. However, expression \eqref{Lig2B} is a first integral of the system \eqref{EQ3.61}, \eqref{EQ3.62}, \eqref{EQ3.63} only if
\begin{align}
    & A \Big[3 A^3 (w_{\text{X}}-1)+A^2 (5 \mu -17) (w_{\text{X}}-1) -A (\mu ( 7 \mu -37)+(\mu -1) (5 \mu -12) w_{\text{X}}+50) \nonumber \\
    & +(\mu -2) (\mu -1) (3 \mu +(\mu -1) w_{\text{X}}-5)\Big]=0. \label{LigNew2B}
\end{align}

Table \ref{tab:R} shows the equilibrium points/sets of the system \eqref{EQ3.61}, \eqref{EQ3.62}, \eqref{EQ3.63} that satisfy the compatibility conditions \eqref{Lig2B} and \eqref{LigNew2B}.
\begin{table}[ht!]
 \caption{Equilibrium points/sets of the system \eqref{EQ3.61}, \eqref{EQ3.62}, \eqref{EQ3.63} that satisfy the compatibility conditions \eqref{Lig2B} and \eqref{LigNew2B}; NH means nonhyperbolic.}
    \label{tab:R}
\resizebox{1\textwidth}{!}{%
 \begin{tabular}{ccccccccc}
 \toprule 
\textbf{Label} & \boldmath{$\Omega_{\text{m}}$} & \boldmath{$\Omega_{\text{X}}$}   & \boldmath{$A$} &  \boldmath{$\mu$} & \boldmath{$q$}  & \boldmath{$w_{\text{X}}$} & \textbf{Stability} \\\midrule
$R_1$& $0$ &$ 0$      &$ \frac{3}{2}$ & $\frac{5}{2}$ & $-\frac{1}{3}$ &$ w_{\text{X}}$ &    Sink for \\
    &&&&&& & $w_{\text{X}}>-1/3$ \\
    &&&&&& & saddle for  $w_{\text{X}}<-1/3$.  \\\midrule
$R_2$& $ \Omega_{\text{m}}$ & $\frac{1}{3} (-3 \Omega_{\text{m}}-2)$    & $\frac{3}{2}$ &$ \frac{7}{2}$ &$-\frac{1}{3} $  & $0$ &  NH. 1D Stable \\
    &&&&&& & manifold \\\midrule
$R_3$& $-\frac{2}{3}$ & $0$    & $\frac{3}{2}$ & $\frac{7}{2}$ & $-\frac{1}{3}$ &$ w_{\text{X}}$  &  NH. 2D Stable \\
    &&&&&& & manifold for $w_{\text{X}}>0$\\
    &&&&&& &  saddle for $w_{\text{X}}<0$ \\\midrule
$R_4$&$ 0$ & $\frac{r+5}{2 (\mu -2)} $  & $\frac{1}{6} (-2 \mu -r+9)$ & $\mu$  &$ -\frac{2 \mu ^2-8 \mu +r+13}{2 (\mu -2)(\mu -1)}$ &$ -\frac{r+7}{4 (\mu -1)} $ & NH.  2 D stable \\  &&&&&&& manifold \\  &&&&&&&  for   $\mu <1\lor \mu >2$ \\
   &&&&&&& Saddle for   $1<\mu<2$ \\\midrule
$R_5$& $0$ &$ -\frac{r-5}{2 (\mu -2)}$  & $\frac{1}{6} (-2 \mu +r+9)$ & $\mu$  & $\frac{-2 \mu ^2+8 \mu +r-13}{2 (\mu -2)(\mu -1)}$ &$\frac{r-7}{4 (\mu -1)}$ & NH. 2D stable \\  &&&&&&& manifold for  $\mu <\frac{7}{2}$ \\
   &&&&&&& saddle for   $\mu >\frac{7}{2}$\\\bottomrule
\end{tabular}}
   
\end{table}

Table \ref{tab:R2} presents the eigenvalues for the equilibrium points/sets of the system \eqref{EQ3.61}, \eqref{EQ3.62}, \eqref{EQ3.63} that satisfy the compatibility conditions \eqref{Lig2B} and \eqref{LigNew2B}. These equilibrium points/sets of the system are enumerated as follows.:
\begin{enumerate}

\item $R_1: \left\{A= \frac{3}{2},\mu = \frac{5}{2},\Omega_{\text{m}}= 0,\Omega_{\text{ X}}= 0\right\}$; $q= -\frac{1}{3}$, and the critical point
\begin{enumerate}
    \item is a sink for $w_{\text{X}}>-1/3$
    \item is a saddle for $w_{\text{X}}<-1/3$.
\end{enumerate}

\item $R_2: \left\{A= \frac{3}{2},\mu = \frac{7}{2},\Omega_{\text{m}}= -\frac{2 }{3},\Omega_{\text{X}}= 0\right\}$; this solution is not physically viable because $\Omega_{\text{m}}<0$.
 The deceleration parameter is $q= -\frac{1}{3}$. It is nonhyperbolic with a one-dimensional stable manifold.

\item $R_3: \left\{A= \frac{3}{2},w_{\text{X}}= 0,\mu = \frac{7}{2},\Omega_{\text{X}}= -\Omega_{\text{m}}-\frac{2}{3}\right\}$. $q= -\frac{1}{3}$; this solution is not physically viable because $\Omega_{\text{X}}<0$
  \begin{enumerate}
    \item is nonhyperbolic with a two-dimensional stable manifold for $w_{\text{X}}>0$
    \item is nonhyperbolic with a one-dimensional stable manifold and a one-dimensional unstable manifold for $w_{\text{X}}<0$ (saddle).
 \end{enumerate}

\item $R_4: \big\{A= \frac{1}{6} \left(-2 \mu -r+9\right),\Omega_{\text{m}}= 0, \Omega_{\text{X}}= \frac{r+5}{2 (\mu-2)}\Big\}$, where \newline $r=\sqrt{8 \mu (2 \mu -9) +105}$. The cosmological parameters are $q=-\frac{2 (\mu -4) \mu +r+13}{2 (\mu -2) (\mu -1)}$ and $w_{\text{X} }= \frac{r+7}{4-4 \mu }$. It is 
\begin{enumerate}
    \item nonhyperbolic with a two-dimensional stable manifold for $\mu <1$, or $\mu >2$
    \item nonhyperbolic with a one-dimensional stable manifold for $1<\mu <2$.
\end{enumerate}
The solution is not physically viable; that is, $\Omega_{\text{X}}\geq 0, A\geq 0$ is not satisfied in the following cases:
\begin{enumerate}
    \item for $\mu>2$, we have $\Omega_{\text{X}}\geq 0$ and $A<0$
    \item for $1\leq \mu <2$, we have $\Omega_{\text{X}}< 0$ and $A\geq 0$
    \item for $\mu<1$, we have $\Omega_{\text{X}}< 0$ and $A< 0$.
\end{enumerate}

\item $R_5: \big\{A= \frac{1}{6} \left(-2 \mu +r+9\right),\Omega_{\text{m}}= 0, \Omega_{\text{X}}= -\frac{r-5}{2 (\mu -2)}\Big\}$, where \newline $ r=\sqrt{8 \mu (2 \mu -9 )+105}$. The cosmological parameters are $q= \frac{-2 (\mu -4) \mu +r-13}{2 (\mu -2) (\mu -1)}$ and $w_{\text{X} }= \frac{r-7}{4 (\mu -1)}$. The equilibrium point is
\begin{enumerate}
    \item nonhyperbolic with a two-dimensional stable manifold for $\mu <\frac{7}{2}$
    \item is a saddle for $\mu >\frac{7}{2}$.
\end{enumerate}
The solution is as follows. 
\begin{enumerate}
\item The solution satisfies $w_{\text{X}}<-1/3$ for $\mu <\frac{5}{2}$
\item The solution is accelerated ($q<0$) for large $t$ for all $\mu$.
\item The solution is not physically viable for $\mu >\frac{5}{2}$.
\end{enumerate}
\end{enumerate}
\begin{table}[ht!]
 \caption{Eigenvalues for the equilibrium points/sets of the system \eqref{EQ3.61}, \eqref{EQ3.62}, \eqref{EQ3.63} that satisfy compatibility conditions \eqref{Lig2B} and \eqref{LigNew2B}.}
    \label{tab:R2}
    \setlength{\tabcolsep}{2.3mm}
 \begin{tabular}{ccccc}
 \toprule 
\textbf{Label} & \boldmath{$\lambda_1$} & \boldmath{$\lambda_2$} & \boldmath{$\lambda_3$}  \\\midrule
$R_1$ & $\frac{15}{2}$ & $-\frac{3}{2}$ & $-\frac{3}{2} (3 w_{\text{X}}+1)$ \\\midrule
$R_2$ &$ -\frac{21}{2}$ & 0 & 0 \\\midrule
$R_3$ & $-\frac{21}{2}$ & 0 &$ -3 w_{\text{X}}$  \\\midrule
$R_4$& 0 & $\frac{1}{6} \left(-16 \mu ^2+72 \mu -2 \mu  r+9 r-105\right)$ & $\frac{1}{6} \left(-12 \mu ^2+61 \mu -3 \mu  r+8 r-84\right)$ \\\midrule
$R_5$ & 0 & $\frac{1}{6} \left(-12 \mu ^2+61 \mu +3 \mu  r-8 r-84\right)$ & $\frac{1}{6} \left(-16 \mu ^2+72 \mu +2 \mu  r-9 r-105\right)$\\\bottomrule 
\end{tabular}
   
\end{table}

\begin{figure}[ht!]
    \includegraphics[scale=1.0]{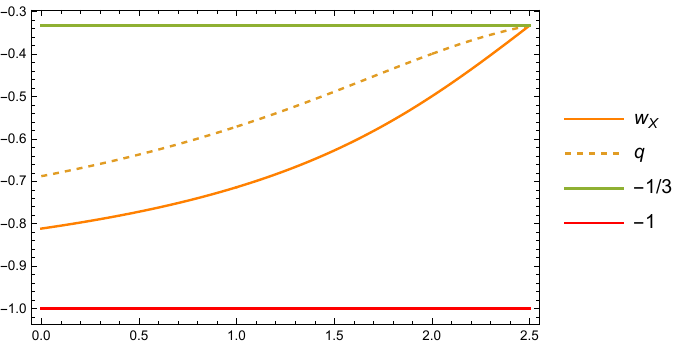}
    \caption{The effective fluid equation of state ($w_{\text{X}}$), with the deceleration parameter ($q$) of the equilibrium point $R_5$ in the physical region in the parameter space $ \mu \in \left(-\infty, \frac{5}{2}\right]$.}
    \label{fig:R_13}
\end{figure}

Figure \ref{fig:R_13} represents the effective state equation of the effective fluid ($w_{\text{X}}$) with the deceleration parameter ($q$) of the equilibrium point $R_{5}$ as a function of $\mu$.

\subsection{Bianchi I Universe} \label{Bianchi}
The line element describes the local and rotationally symmetric Bianchi I spacetime:
\begin{equation}
ds^{2}=-N^{2}\left( t\right) dt^{2}+e^{2\alpha\left( t\right) }\left(e^{2\beta\left ( t\right)}dx^{2}+e^{-\beta\left( t\right) }\left(
dy^{2}+dz^{2}\right) \right), \label{ch.03}%
\end{equation}
where $\alpha(t)$ is the scale factor for the three-dimensional hypersurface, $\beta(t)$ is the anisotropic parameter, and $N(t)$ is the lapse function. The metric (\ref{ch.03}) reduces to the spatially flat FLRW geometry in the limit $\beta \rightarrow0$. 
The effective fractional action can be written as

\begin{align}
& S_{\text{eff}} = \frac{1}{\Gamma(\alpha)}\int_0^t \Bigg[\frac{3}{8\pi G N^2(\theta)}\Bigg( \ddot{\alpha}(\theta)-\frac{\dot{\alpha}(\theta)\dot{N}(\theta)}{N(\theta)} 
+2\dot{\alpha}^{2}(\theta)+\frac{1}{4}\dot{\beta}^{2}(\theta)\Bigg)  +e^{3\alpha(\theta)}\mathcal{L}_{\text{m}}\Bigg] (t-\theta)^{\mu-1} N(\theta) d\theta. \label{LFCBI}
\end{align}

Varying the Action \eqref{LFCBI} for $q_i\in \{N,\alpha, \beta\}$ and assuming the lapse function $N=1$ after the variation, from the Euler--Poisson Equations \eqref{EP} we obtain 
\begin{align}
   & \dot{\alpha}^2 +\frac{(1-\mu) \dot{\alpha} }{t} -\frac{1}{4} \dot{\beta}^2 = \frac{ 8 \pi G}{3} \rho,
\\
   &\ddot{\alpha } +\frac{(1-\mu ) \dot{\alpha} }{t}+\frac{3}{2}{\dot{\alpha}}^2 +\frac{ 3}{8} \dot{\beta}^2 +\frac{(\mu -2) (\mu -1)}{2 t^2}=-4 \pi G p,
\\
   & \dot{\beta} \left(3 \dot{\alpha} +\frac{1-\mu }{t}\right)+\ddot{\beta} =0,
\end{align}
where $\rho= \sum_i\rho_i$ and $p =\sum_i p_i$ denote the total energy density and total pressure of the matter fields, respectively, and $H=\dot{\alpha}$ and $\sigma=\dot{\beta}/2$ are the respective Hubble and anisotropy parameters. To designate the temporary independent variables, we use the rule $t-\theta \mapsto t$, $\theta \mapsto t$ \cite{Shchigolev:2010vh}, where the dots denote these derivatives.

Therefore, the field equations can be written as
\begin{align}
   & H^2 +\frac{(1-\mu) H}{t} -\sigma^2= \frac{8 \pi G}{3} \sum_i\rho_i, \label{BIb}\\
   & \dot{H}+\frac{(1-\mu) H}{t}+\frac{3}{2}{H}^2 + \frac{(\mu -2) (\mu -1 )}{2 t^2}+\frac{3}{2} \sigma^2 = - 4 \pi G \sum_i p_i, \label{BIa}\\
   & \dot{\sigma}+ 3\sigma \left(H+\frac{1-\mu }{3 t}\right)=0,
\end{align}
and we separately assume conservation equations
\begin{equation}
     \dot{\rho}_i+3\left(H+\frac{1-\mu}{3t}\right)(\rho_i+p_i)=0. \label{cons2}
\end{equation}

For $\alpha \neq 1$, \eqref{cons2} produces modified continuity equations only if
\begin{align}
    \frac{8 \pi G}{3} \sum_i p_i=\frac{2 (\mu -3) H}{t}+H^2-\frac{(\mu -2) (\mu -1)}{t^2}-\sigma^2.\label{BIc}
\end{align}

Removing $\sum_i p_i$ and $\sum_i \rho_i$ from \eqref{BIb}, \eqref{BIa}, and \eqref{BIc} results in the cancellation of the $\sigma$ terms in the Equation \eqref{BIa}. Therefore, the master Equation \eqref{NewIntegrableH} is obtained, which has a solution \eqref{solution} where $c_1$ is an integration constant that depends on $\mu$, the value of $H$ today, $H_0 $, and the age of the universe $t_U$. This leads to $a(t)\simeq t^{\frac{1}{6} \left(-2 \mu +r+9\right )}$ for large $t$, whence we acquire $q<0$, resulting in late-time acceleration without dark energy.

\subsubsection{Dynamical Systems and Stability Analysis}
\label{AlternattiveII}
Continuing with our analysis, we assume that the matter components have the equation of state  $p_i=w_i\rho_i$, where $w_i \neq -1 $ are constants \cite{Garcia-Aspeitia:2022uxz}. Then, using the dimensionless variables \eqref{AlternativevarsFLRW} and
\begin{equation} \Sigma= {\sigma}/{H} \label{AlternativevarsBI}
\end{equation}
and with the new derivative $f^{\prime}=\dot{f}/H$, for $\mu \neq 1$ we obtain
\begin{align}
\Omega_j^{\prime}& =\Omega_j \left[
   (2 q-3 w_j-1) + (w_j+1) \left(1- \Sigma^2 - \sum_i\Omega_i\right)\right],\\
  \Sigma^{\prime} &=\Sigma \left[ (q-2) + \left(1- \Sigma^2 - \sum_i\Omega_i\right)\right], 
\end{align}  
and the age parameter evolves according to \eqref{Evol-Age}, where $q$ is found based on Equation \eqref{NewIntegrableH} (valid for the FLRW and Bianchi I metrics) as
\eqref{decc}.

Instead of following \cite{Garcia-Aspeitia:2022uxz}, we present an alternative approach using the Riccati Equation \eqref{NewIntegrableH} while assuming that the matter components have equations of state $p_i=w_i\rho_i$, where $w_i \neq -1$ are constants. 

It can be observed that Equations \eqref{BIb} and \eqref{BIc} impose restrictions on the type of matter components of the universe; using, say, the dimensionless variables \eqref{AlternativevarsFLRW} and \eqref{AlternativevarsBI}, we obtain
\begin{align}
   1+\frac{(1-\mu)}{A} -\Sigma^2 &=\sum_i\Omega_i,
\\
1+ \frac{2 (\mu -3)}{A} -\frac{ (\mu -2) (\mu -1)}{A^2} -\Sigma^2 &= \sum_i w_i \Omega_i.
\end{align}

Compared with the standard model, it is assumed that the components of the Universe are   CDM   and radiation. Then, we have the constraints
\begin{align}
 1+\frac{(1-\mu)}{A} -\Sigma^2 &=\Omega_{\text{m}} +\Omega_{\text{r}}, \label{LigC}
\\
1+ \frac{2 (\mu -3)}{A}-\frac{ (\mu -2) (\mu -1)}{A^2} -\Sigma^2& = \frac{1}{ 3} \Omega_{\text{r}}. \label{LigD}
\end{align}

The dimensionless energy densities evolve according to
\begin{align}
\Omega_{\text{m}}^{\prime}& =\Omega_{\text{m}} (2 q-\Omega_{\text{m}}-\Omega_{\text{r}}-\Sigma^2),\\
\Omega_{\text{r}}^{\prime}& =\frac{2}{3} \Omega_{\text{r}} (3 q-2 \Omega_{\text{m}}-2 \Omega_{\text{r}}-2\Sigma^2-1),\\
  \Sigma^{\prime} &=\Sigma \left( q-1 -\Omega_{\text{m}}-\Omega_{\text{r}}-\Sigma^2 \right), 
\end{align}
and the age parameter evolves according to \eqref{Evol-Age}, where $q$ is provided by Equation \eqref{decc} and  $f^{\prime}=\dot{f}/H$.

Using the derivative \eqref{NEW-TIME},
the following dynamical system is obtained:
\begin{align}
\frac{d \Omega_{\text{m}}}{d \tau}&=\Omega_{\text{m}}\left(5 A \mu +A (3 A-17)-2 \mu ^2+6 \mu -4\right), \label{EQ3.86}\\
\frac{d \Omega_{\text{r}}}{d \tau}&=\frac{2}{3} \Omega_{\text{r}}\left(8 A \mu +A (3 A-26)-3 \mu ^2+9 \mu -6\right), \label{EQ3.87}\\
\frac{d \Sigma}{d \tau}&=\Sigma  (3 A (\mu -3)-(\mu -2) (\mu -1)), \label{EQ3.88}\\
\frac{d A}{d \tau}&=A \left(-2 A \mu -3 (A-3) A+\mu   ^2-3 \mu +2\right),  \label{EQ3.89}
\end{align}
with $\left(1- \Sigma^2- \sum_i\Omega_i\right)=(1-\mu)A^{-1}$ used to obtain decoupled equations.

The expression \eqref{LigC} is trivially a first integral of the system \eqref{EQ3.86}, \eqref{EQ3.87}, \eqref{EQ3.88}, \eqref{EQ3.89}. 
However, the expression \eqref{LigD} is a first integral of the system \eqref{EQ3.86}, \eqref{EQ3.87}, \eqref{EQ3.88}, \eqref{EQ3.89} only if

\begin{align}
 & A \Big[3 A^3 \left(\Sigma ^2-1\right)+A^2 \left(-\mu \left(\Sigma ^2+5\right)+\Sigma ^2+17\right)+A ((64-13 \mu )
   \mu -81) \nonumber\\
   & +(\mu -2) (\mu -1) (5 \mu -8)\big]=0. \label{LigNewC}
\end{align}

The equilibrium points of the system \eqref{EQ3.86}, \eqref{EQ3.87}, \eqref{EQ3.88}, \eqref{EQ3.89} that satisfy the compatibility conditions \eqref{LigC} and \eqref{LigNewC} and $A\neq 0$ are presented in Table \ref{tab:S}. 
\begin{table}[ht!]
 \caption{Equilibrium points of the system \eqref{EQ3.86}, \eqref{EQ3.87}, \eqref{EQ3.88}, \eqref{EQ3.89} that satisfy the compatibility conditions \eqref{LigC} and \eqref{LigNewC}. $\lambda_1, \lambda_2, \lambda_3$ and $\lambda_4$ are the eigenvalues; NH means nonhyperbolic. }
    \label{tab:S}
     \setlength{\tabcolsep}{2.7mm}
 \begin{tabular}{cccccccccccc}
 \toprule 
\textbf{Label} & \boldmath{$\Omega_{\text{m}}$} & \boldmath{$\Omega_{\text{r}}$} & \boldmath{$\Sigma$} & \boldmath{$A$} &  \boldmath{$\mu$} & \boldmath{$q$} & \boldmath{$\lambda_1$} & \boldmath{$\lambda_2$} & \boldmath{$\lambda_3$}&  \boldmath{$\lambda_4$} & \textbf{Stability} \\\midrule
 $S_1$ & $0$ & $0$ &$ 0$ & $\frac{3}{2}$ & $\frac{5}{2}$ & $-\frac{1}{3}$ & $-\frac{15}{2}$ & $-3$ & $-3 $& $-\frac{3}{2}$ & Sink \\ \midrule
 $S_2$ & $ -\frac{2}{3}$ & $0$ &$ 0$ &$ \frac{3}{2}$ & $\frac{7}{2}$ &$ -\frac{1}{3}$ & $-\frac{21}{2}$ & $-\frac{3}{2}$ &$-1$ &$ 0$ & NH \\ \midrule
 $S_3$ & $ 0 $&$ -\frac{6}{5}$ & $0$ & $\frac{15}{8}$ & $\frac{41}{8}$ & $-\frac{7}{15}$ &$ -\frac{375}{16} $&$ -\frac{15}{16}$ & $ \frac{15}{16}$ &$ 0 $ & NH \\\bottomrule 
\end{tabular}
   
\end{table}

The results are listed below:
\begin{enumerate}
    \item $S_1: \left\{A= \frac{3}{2}, \Omega_{\text{m}} = 0, \Omega_{\text{r}} = 0,\Sigma = 0,\mu = \frac{5}{2}\right\}$. As the deceleration parameter is $q=-\frac{1}{3}$, the solution is accelerated. It is a sink.
    \item $S_2: \left\{A= \frac{3}{2}, \Omega_{\text{m}}= -\frac{2}{3}, \Omega_{\text{r}} = 0,\Sigma = 0, \mu = \frac{7}{2}\right\}$. It is not physically viable  because $\Omega_{\text{m}}<0$.
    \item $S_3: \left\{A= \frac{15}{8}, \Sigma = 0, \mu = \frac{41}{8}\right\}$. It is not physically viable  because $\Omega_{\text{r}}<0$.
\end{enumerate}

Then, based on the compatibility conditions of the problem, considering that the matter sources in the Bianchi I metric are radiation and cold dark matter, and assuming $\mu \notin\{1,2\}$, it is imposed that the parameter $\mu$ can take discrete values
$\{5/2, 7/2, 41/8\}$.

A second alternative formulation is the following: the equation of state of one of the matter sources is not imposed; instead, it is deduced from the compatibility conditions.
That is, it is assumed that the components of the universe are  CDM and a fluid with a constant state equation to be determined.
Then, we have the constraints
\begin{align}
 1+\frac{(1-\mu)}{A}  &=\Omega_{\text{m}} +\Omega_{\text{X}} + \Sigma^2, \label{Lig3A}
\\
1+  \frac{2 (\mu -3)}{A}-\frac{ (\mu -2) (\mu -1)}{A^2}& = w_{\text{X}}  \Omega_{\text{X}} + \Sigma^2. \label{Lig3B}
\end{align}

The dimensionless energy densities  evolve according to
\begin{align}
\Omega_{\text{m}}^{\prime}& =\Omega_{\text{m}} (2 q-\Omega_{\text{m}}-\Omega_{\text{X}}-\Sigma^2),\\
\Omega_{\text{X}}^{\prime}& =\Omega_{\text{X}} \left[ (2 q-3 w_{\text{X}} -1) +  (w_{\text{X}} +1)   \left(1 -\Omega_{\text{m}}-\Omega_{\text{X}}-\Sigma^2\right)\right],\\
\Sigma^{\prime} &=\Sigma \left[ q-1 - \Omega_{\text{m}}-\Omega_{\text{X}} -\Sigma^2\right], 
\end{align}
the age parameter evolves according to  \eqref{Evol-Age},  and  $f^{\prime}=\dot{f}/H$. 

Using the derivative \eqref{NEW-TIME},
the following dynamical system is obtained:
\begin{align}
\frac{d \Omega_{\text{m}}}{d \tau}&= \Omega_{\text{m}} \left(5 A \mu +A (3 A-17)-2 \mu ^2+6 \mu -4\right), \label{EQ3.113}\\
\frac{d \Omega_{\text{X}}}{d \tau}&=\Omega_{\text{X}} \left[A \mu (w_{\text{X}}+5)-A (3 A (w_{\text{X}}-1)+w_{\text{X}}+17)-2 \mu ^2+6 \mu -4\right], \label{EQ3.114}\\
\frac{d \Sigma}{d \tau}&=\Sigma  (3 A (\mu -3)-(\mu -2) (\mu  -1)), \label{EQ3.115}
\\
\frac{d A}{d \tau}&= A \left(-2 A \mu -3 (A-3) A+\mu ^2-3 \mu +2\right),  \label{EQ3.116}
\end{align}
with the relation $\left(1- \Sigma^2- \sum_i\Omega_i\right)=(1-\mu)A^{-1}$ used to obtain decoupled equations.

The expression \eqref{Lig3A} is trivially a first integral of the system \eqref{EQ3.113}, \eqref{EQ3.114}, \eqref{EQ3.115}, \eqref{EQ3.116}. However, the expression \eqref{Lig3B} is a first integral of the system \eqref{EQ3.113}, \eqref{EQ3.114}, \eqref{EQ3.115}, \eqref{EQ3.116} only if
\begin{small}
\begin{align}
    & A \Big[-3 A^3 \left(\Sigma ^2-1\right) (w_{\text{X}}-1)+A^2 (w_{\text{X}}-1) \left((\mu -1) \Sigma ^2+5 \mu -17\right) \nonumber \\
    & -A    (\mu  (7 \mu -37)+(\mu -1) (5 \mu -12) w_{\text{X}}+50)+(\mu -2) (\mu -1) (3 \mu +(\mu -1) w_{\text{X}}-5)\Big]=0. \label{LigNew3B}
\end{align}
\end{small}

Table \ref{tab:S2} shows the equilibrium points/sets of the system \eqref{EQ3.113}, \eqref{EQ3.114}, \eqref{EQ3.115}, \eqref{EQ3.116} that satisfy the compatibility conditions \eqref{Lig3B} and \eqref{LigNew3B}. The analysis of the equilibrium points with $A=0$ is omitted here due to physical considerations. 
\begin{table}[ht!]
  \caption{Equilibrium points/sets of the system \eqref{EQ3.113}, \eqref{EQ3.114}, \eqref{EQ3.115}, \eqref{EQ3.116} that satisfy the compatibility conditions \eqref{Lig3B} and \eqref{LigNew3B}; NH means nonhyperbolic.}
    \label{tab:S2}
\resizebox{1\textwidth}{!}{%
 \begin{tabular}{ccccccccc}
 \toprule 
\textbf{Label} & \boldmath{$\Omega_{\text{m}}$} & \boldmath{$\Omega_{\text{X}}$} & \boldmath{$\Sigma$} & \boldmath{$A$} &  \boldmath{$\mu$} & \boldmath{$q$} & \boldmath{$w_{\text{X}}$}  & \textbf{Stability}\\ \midrule
  $R_1$&$0$ & $0$ & $0$& $\frac{3}{2}$ & $\frac{5}{2}$ & $-\frac{1}{3}$ &$ w_{\text{X}}$  & Sink for $w_{\text{X}}>-\frac{1}{3}$\\ 
  &&&&&&&& Saddle for  $ w_{\text{X}}<-\frac{1}{3}$\\  \midrule
  $R_2$& $ \Omega_{\text{m}}$ & $\frac{1}{3} (-3 \Omega_{\text{m}}-2)$ & 0 & $\frac{3}{2}$ & $\frac{7}{2}$ &$ -\frac{1}{3}$ & $0$ & NH \\ 
    &&&&&&&& 2D stable \\ 
    &&&&&&&& manifold \\\midrule
  $R_3$&$ -\frac{2}{3}$ &$0$ & $0$ &$ \frac{3}{2}$ &$ \frac{7}{2}$ &$ -\frac{1}{3}$ & $w_{\text{X}}$ & NH \\ 
    &&&&&&&& 3 D stable \\ 
    &&&&&&&& manifold   for $w_{\text{X}}>0$ \\
    &&&&&&&& Saddle for  $w_{\text{X}}<0$ \\\midrule
  $R_4$&$0$&$ \frac{r+5}{2 (\mu -2)}$ &$ 0$ &$ \frac{1}{6} (-2 \mu -r+9)$ & $\mu $ & $-\frac{2 \mu ^2-8 \mu +r+13}{2 \mu ^2-6 \mu +4}$ & $\frac{r+7}{4-4 \mu }$  & NH.  3D stable \\  &&&&&&&& manifold\\  &&&&&&&& for   $\mu <1\lor \mu >2$ \\
   &&&&&&&& Saddle for  $1<\mu<2$ \\\midrule
  $R_5$&$0$& $-\frac{r-5}{2 (\mu -2)}$ & $0$ & $\frac{1}{6} (-2 \mu +r+9) $& $\mu$  & $\frac{-2 \mu ^2+8 \mu +r-13}{2 \left(\mu ^2-3 \mu +2\right)}$ & $\frac{r-7}{4 (\mu -1)}$ & NH.  3D stable \\  &&&&&&&& manifold \\  &&&&&&&& for   $\mu <\frac{7}{2}$ \\
   &&&&&&&& saddle for   $\mu >\frac{7}{2}$ \\\bottomrule
\end{tabular}}
  
\end{table}

The corresponding eigenvalues are presented in Table \ref{tab:S2b}.
Below are the equilibrium points/sets of the system \eqref{EQ3.113}, \eqref{EQ3.114}, \eqref{EQ3.115}, \eqref{EQ3.116} that satisfy the compatibility conditions \eqref{Lig3B} and \eqref{LigNew3B} and $A\neq 0$. 
\begin{enumerate}

\item $R_1: \left\{A= \frac{3}{2},\mu = \frac{5}{2},\Sigma = 0,\Omega_{\text{m}}= 0,\Omega_{\text{X}}= 0\right\}$. As the deceleration parameter is $q=-\frac{1}{3}$, the solution is accelerated. It is 
\begin{enumerate}
\item a sink for $ w_{\text{X}}>-\frac{1}{3}$
\item a saddle for $ w_{\text{X}}<-\frac{1}{3}$.
\end{enumerate}

\item $R_2: \left\{A= \frac{3}{2},w_{\text{X}}= 0,\mu = \frac{7}{2},\Sigma = 0 ,  \Omega_{\text{X}}=\frac{1}{3} (-3 \Omega_{\text{m}}-2)\right\}$. It is nonhyperbolic with a two-dimensional stable manifold. This solution is not physically viable because $\Omega_{\text{X}}<0$.

\item $R_3: \left\{A= \frac{3}{2},\mu = \frac{7}{2},\Sigma = 0,\Omega_{\text{m}}= -\frac{2}{3},\Omega_{\text{X}}= 0\right\}$. This solution is not physically viable because $\Omega_{\text{m}}<0$.
 As deceleration parameter is $q= -\frac{1}{3}$, the solution is accelerated. It is  
\begin{enumerate}
    \item nonhyperbolic with a three-dimensional stable manifold for $w_{\text{X}}>0$
    \item is nonhyperbolic with a two-dimensional stable manifold and a one-dimensional unstable manifold for $w_{\text{X}}<0$ (saddle).
 \end{enumerate}
\item $R_4: \left\{\Omega_{\text{m}}= 0,\Omega_{\text{X}}= \frac{r+5}{2 (\mu -2)} ,\Sigma = 0,A= \frac{1}{6} \left(-2 \mu -r+9\right)\right\}$. The cosmological parameters are $q=-\frac{2 (\mu -4) \mu +r+13}{2 (\mu -2) (\mu -1)}$ and $w_{\text{X} }= \frac{r+7}{4-4 \mu }$.
The equilibrium point is
\begin{enumerate}
    \item nonhyperbolic with a three-dimensional stable manifold for $\mu <1$, or $\mu >2$
    \item a nonhyperbolic saddle for $1<\mu <2$.
\end{enumerate}

The solution is not physically viable; that is, $\Omega_{\text{X}}\geq 0, A\geq 0$ is not satisfied in the following cases:
\begin{enumerate}
    \item for $\mu>2$, we have $\Omega_{\text{X}}\geq 0$ and $A<0$
    \item for $1\leq \mu <2$, we have $\Omega_{\text{X}}< 0$ and $A\geq 0$
    \item for $\mu<1$, we have $\Omega_{\text{X}}< 0$ and $A< 0$.
\end{enumerate}

\item $R_5: \left\{\Omega_{\text{m}}= 0,\Omega_{\text{X}}= -\frac{r-5}{2 (\mu -2) },\Sigma = 0,A= \frac{1}{6}
   \left(-2 \mu +r+9\right)\right\}$. The cosmological parameters are $q= \frac{-2 (\mu -4) \mu +r-13}{2 (\mu -2) (\mu -1)}$ and $w_{\text{X} }= \frac{r-7}{4 (\mu -1)}$.
The equilibrium point is
\begin{enumerate}
    \item nonhyperbolic with a three-dimensional stable manifold for $\mu <\frac{7}{2}$
    \item is a saddle for $\mu >\frac{7}{2}$.
\end{enumerate}
The solution is as follows:
\begin{enumerate}
\item the solution satisfies $w_{\text{X}}<-1/3$ for $\mu <\frac{5}{2}$
\item the solution is always accelerated ($q<0$) for large $t$; anisotropy decays very fast and does not influence late-time behaviour
\item the solution is not physically viable for $\mu >\frac{5}{2}$.
\end{enumerate}
\end{enumerate}

\begin{table}[ht!]
 \caption{Eigenvalues for the equilibrium points/sets of the system \eqref{EQ3.113}, \eqref{EQ3.114}, \eqref{EQ3.115}, \eqref{EQ3.116} that satisfy the compatibility conditions \eqref{Lig3B} and \eqref{LigNew3B}. }
    \label{tab:S2b}
\resizebox{1\textwidth}{!}{%
 \begin{tabular}{ccccc}
 \toprule 
 \textbf{Label}&  \boldmath{$\lambda_1$} & \boldmath{$\lambda_2$} & \boldmath{$\lambda_3$}&  \boldmath{$\lambda_4$} \\\midrule
 $R_1$& $-\frac{15}{2}$ & $ -3$ & $-\frac{3}{2} $& $-\frac{3}{2}  (3 w_{\text{X}}+1)$ \\ \midrule
 $R_2$ & $ -\frac{21}{2}$ & $-\frac{3}{2}$ & $0$ & $0$ \\ \midrule
 $R_3$ & $-\frac{21}{2}$ & $-\frac{3}{2}$ & $0$ & $-3
   w_{\text{X}}$ \\ \midrule
 $R_4$ & $0 $& $\frac{1}{2} (-\mu  (4 \mu +r-21)+3 r-31)$ & $\frac{1}{6} (-2 \mu  (8 \mu +r-36)+9
   r-105)$ & $\frac{1}{6} (\mu  (-12 \mu -3 r+61)+8 r-84)$ \\ \midrule
 $R_5$ & $0$ & $\frac{1}{6} (\mu  (-12 \mu +3 r+61)-8 r-84)$ & $\frac{1}{6} (2 \mu 
   (-8 \mu +r+36)-3 (3 r+35)) $& $\frac{1}{2} (\mu  (-4 \mu +r+21)-3 r-31)$ \\\bottomrule
\end{tabular}}
   
\end{table}
\section{Fractional Formulation of Cosmology with Scalar Field and Matter}\label{ch_4}

This section discusses a fractional formulation of cosmology with scalar field and matter.

\subsection{Flat FLRW Model}

The fractional action can be written as
\begin{align}
     S_{EH}^\mu&=\frac{1}{\Gamma(\mu)}\int_{0}^{t}N\left[\frac{3}{8\pi G}\left(\frac {a^2(\theta)\Ddot{a}(\theta)}{N^2(\theta)}+\frac{a(\theta)\Dot{a}^2(\theta)}{N^2(\theta)}-\frac{a^2(\theta)\Dot{a}(\theta)\Dot{N}(\theta) }{N^3(\theta)}-\frac{\Lambda a^3(\theta)}{3}\right) \right. \nonumber\\
     & \;\;\;\;\;\;\;\;\;\;\;\;\;\;\;\;\;\;\;\;\;\;\;\;\;\; \left.+a^3(\theta)\left(\frac{\epsilon \Dot{\phi}^2(\theta)}{2N^2(\theta)}-V(\phi(\theta))\right)\right](t-\theta)^{\mu-1}d\theta, \label{[ACTION]}
\end{align}
where $a$ is the scale factor, $G$ is Newton's universal gravitation constant, $N$ is a lapse function that is equal to one after performing the action variation,  $\Lambda$ is the cosmological constant, $\phi$ is the scalar field, $V$ is the potential of the scalar field, $\epsilon$ is a constant that can be $\pm 1$, as we can have a scalar field with positive (quintessence) or negative ({phantom}) kinetic energy, $t$ is the cosmic time, and $\theta$ is the proper time of the system.

By varying the action \eqref{[ACTION]} with respect to $\{\phi, a, N\}$ and making the replacement $N=1$ after the variation, the following equations are obtained:
\begin{equation}
     \Ddot{\phi}+3\left(H+\frac{1-\mu}{3t}\right)\Dot{\phi}+\epsilon\frac{dV(\phi)}{d\phi}= 0,
     \label{phi:}
\end{equation}
\begin{equation}
     \Dot{H}-\frac{1-\mu}{2t}H+\frac{(1-\mu)(2-\mu)}{2t^2}=-4\pi G \left(\epsilon \Dot{\phi}^2+\rho + p\right),
\end{equation}
\begin{equation}
     H^2+\frac{1-\mu}{t}H=\frac{8\pi G}{3}\left(\epsilon \frac{\Dot{\phi}^2}{2}+V (\phi)+\rho \right)+\frac{\Lambda}{3}.
\label{H^2}\end{equation} 
where $\rho=\sum_{i} \rho_i$ is the energy density of all matter sources other than the scalar field and $p=\sum_{i} p_i$ is the corresponding pressure.
To designate the temporary independent variables, the rule $t-\theta \mapsto t$, $\theta \mapsto t$ \cite{Shchigolev:2010vh} is used, with the dots denoting these derivatives. 

Assuming that all matter sources with density $\rho_i$ are separately conserved, we can rewrite the conservation equation for the $i$th component as
\begin{equation}
      \Dot{\rho_i}+3\left(H+\frac{1-\mu}{3t}\right)(\rho_i+p_i)=0. \label{EQ-CNS}
\end{equation}

To ensure that Friedmann's formula is preserved, that is, \eqref{H^2} is a first integral of the system, and that \eqref{EQ-CNS} is satisfied, the following equations are deduced:
\begin{align}
   &\sum_i \rho_i= -\frac{3 (\mu -1) H}{8 \pi G t}+\frac{3 H^2}{8 \pi G}-\frac{\Lambda }{8 \pi G}-V(\phi)-\frac{1}{2} \epsilon {\dot{\phi }}^2, \\
   & \sum_i p_i= \frac{3 (\mu -3) H}{4 \pi G t}+\frac{3 H^2}{8 \pi G}+\frac{\Lambda t^2-3 (\mu -2) (\mu -1)}{8 \pi G t^2}+V(\phi)-\frac{1}{2} \epsilon {\dot{\phi }}^2.
\end{align}

These equations allow an effective state equation for matter to be deduced without imposing state equations on each matter fluid. 
The equations of motion for the other matter sources are decoupled, allowing the simplified system to be investigated.
\begin{align}
   & \dot{H}= \frac{2(4- \mu ) H}{t}-3 H ^2+\frac{(\mu -2) (\mu -1)}{t^2},\\
   & \ddot{\phi}= -3 H  \dot{\phi}+\frac{(\mu -1) \dot{\phi}}{t}-\epsilon  V'(\phi).
\end{align}

Note that for $\mu \neq 1$, the first equation is Riccati's ordinary differential Equation \eqref{NewIntegrableH}, for which the analytic solution is \eqref{solution}, where $c_1$ is an integration constant defined by \eqref{eq:c1_newH} which depends on $\mu$, the value $H_0$, and the age of the universe $t_U$. As before, for large $t$ the asymptotic scale factor can be expressed as $a(t)\simeq t^{\frac{1}{6} \left(-2 \mu +r+9\right)}$,
allowing the acceleration of the late universe to be obtained.
These results are independent of the matter source and anisotropy.

\subsubsection{Analysis of Dynamical Systems}
\label{Sf_First.System}

In this section, it is assumed that the potential is exponential, that is,
\begin{equation}
      V(\phi)=V_0 e^{-4 \sqrt{3 \pi } \sqrt{G} \lambda \phi },
\end{equation} where we assume that $\lambda>0$ and add only one matter source, with density $\rho_m$ and pressure $p_m$.

Then, the following variables are defined:
\begin{align}
      x= \frac{\sqrt{8 \pi G}}{\sqrt{6}H} \dot{\phi}, \; y= \frac{1}{H}\sqrt{\frac{8 \pi G V(\phi)}{3}}, \; A=tH, \; \Omega_\Lambda = \frac{\Lambda}{3 H^2}, \; \Omega_m= \frac{8 \pi G \rho_m}{3 H^2}, \label{sf-vars}
\end{align}
satisfying the constraint
\begin{equation}
      1 +\frac{1-\mu}{A} = \epsilon x^2 + y^2 + \Omega_\Lambda + \Omega_m,
\end{equation}
which is used as the definition of $\Omega_m$.
With the new derivative $f^{\prime}=\dot{f}/H$, we obtain the following dynamical system for $\mu \neq 1$:
\begin{align}
      & x^{\prime}= (q-2) x + \frac{(\mu -1) x}{A}+ 3 \lambda y^2 \epsilon,
\\
      & y^{\prime} = y (q-3 \lambda x+1),
\\
      & \Omega_\Lambda^{\prime}=2\Omega_\Lambda(1+q),
\end{align} 
and  \eqref{Evol-Age}, where $q$ is defined by \eqref{decc}. 

Hence,
\begin{align}
   x^{\prime}&= -\frac{(\mu -2) (\mu -1) x}{A^2}+\frac{3 (\mu -3) x}{A}+3 \lambda y^2 \epsilon,
\\
     y^{\prime}&= -\frac{(\mu -2) (\mu -1) y}{A^2}+\frac{2 (\mu -4) y}{A}+y ( 3-3 \lambda x),
\\
    A^{\prime}&= \frac{(\mu -2) (\mu -1)}{A}-3 A-2 \mu+9,
\\
   \Omega_\Lambda^{\prime} &= \Omega_\Lambda \left(-\frac{2 (\mu -2) (\mu -1)}{A^2}+\frac{4 (\mu - 4)}{A}+6\right). \label{eqLambda}
\end{align}

Note that Equation \eqref{eqLambda} is decoupled, thereby obtaining a reduced system for $(x,y, A)$ with $A\neq 0$.
Using the time variable \eqref{NEW-TIME},
the following dynamical system is obtained:
\begin{align}
  & \frac{d x}{d\tau}=  3 A^2 \lambda  y^2 \epsilon +3 A (\mu -3) x-(\mu -2) (\mu -1) x, \label{s1}\\
   & \frac{d y}{d\tau}= A^2 y (3-3 \lambda  x)+2 A (\mu -4) y-(\mu -2)   (\mu -1) y, \label{s2}\\
     & \frac{d A}{d\tau}= -3 A^3+A^2 (9-2 \mu )+A (\mu -2) (\mu -1), \label{s3}
\end{align}
which is defined on the phase space
\begin{equation}
   A(1-  \epsilon x^2- y^2)  +(1-\mu):= A(\Omega_m +\Omega_\Lambda)\geq 0.
\end{equation}

\begin{table}[ht!]
\caption{Equilibrium points of the dynamical system \eqref{s1}, \eqref{s2}, \eqref{s3} with $A\neq 0$, where $r=\sqrt{8\mu(2\mu-9)+105}$.}
\label{tab:critical points}
\setlength{\tabcolsep}{6.15mm}
\begin{tabular}{cccc}
\toprule
\textbf{Labels} & \boldmath{$x$} & \boldmath{$y$} & \boldmath{$A$}\\
\midrule
$Sf_1$ & $-\frac{2}{\lambda  \left(2 \mu +r-9\right)}$    & $-\frac{\sqrt{-\left(\left(4 \mu
   +r-9\right) \mu ^2\right)-6 \left(r+9\right)}}{3 \sqrt{2}
   \lambda  \left(\mu ^2-3 \mu +2\right) \sqrt{\epsilon }}$ & $\frac{1}{6} \left(-2 \mu -r+9\right)$  \\ [0.9ex]
\midrule
$Sf_2$ & $ -\frac{2}{\lambda  \left(2 \mu +r-9\right)}$    & $ \frac{\sqrt{-\left(\left(4 \mu
   +r-9\right) \mu ^2\right)-6 \left(r+9\right)}}{3 \sqrt{2}
   \lambda  \left(\mu ^2-3 \mu +2\right) \sqrt{\epsilon }}$ & $ \frac{1}{6} \left(-2 \mu -r+9\right)$  \\ [0.9ex]
\midrule
$Sf_3$ & $0$    & $0$ & $\frac{1}{6} \left(-2 \mu -r+9\right)$  \\ [0.9ex]
\midrule
$Sf_4$ & $\frac{2}{\lambda  \left(-2 \mu +r+9\right)}$    & $-\frac{\sqrt{\left(-4 \mu
   +r+9\right) \mu ^2+6 \left(r-9\right)}}{3 \sqrt{2} \lambda 
   \left(\mu ^2-3 \mu +2\right) \sqrt{\epsilon }}$ & $ \frac{1}{6} \left(-2 \mu +r+9\right)$  \\ [0.9ex]
\midrule
$Sf_5$ & $\frac{2}{\lambda  \left(-2 \mu +r+9\right)}$    & $ \frac{\sqrt{\left(-4 \mu +r+9\right) \mu ^2+6 \left(r-9\right)}}{3 \sqrt{2} \lambda  \left(\mu
   ^2-3 \mu +2\right) \sqrt{\epsilon }}$ & $ \frac{1}{6} \left(-2 \mu +r+9\right)$  \\ [0.9ex]
\midrule
$Sf_6$ & $0$    & $0$ & $\frac{1}{6} \left(-2 \mu +r+9\right)$  \\ [0.9ex]
\bottomrule
\end{tabular}
\end{table}

Table \ref{tab:critical points} presents the equilibrium points of the dynamical system \eqref{s1}, \eqref{s2}, \eqref{s3} with $A\neq 0$, where $r=\sqrt{8\mu(2\mu-9)+105}$.

\subsubsection{Stability Analysis for $\epsilon=1$}
The equilibrium points of the system \eqref{s1}, \eqref{s2}, \eqref{s3} for $\epsilon=1$ with $A\neq 0$ are:
\begin{enumerate}
      \item $Sf_1: \big(
     -\frac{2}{\lambda (2\mu+r-9)},
     -\frac{\sqrt{-\frac{1}{2} \mu ^2 (4 \mu +r-9)-3
     (r+9)}}{3 \lambda \left(\mu ^2-3 \mu
     +2\right)}, \frac{1}{6} (-2 \mu
     -r+9)\big),$ where \newline $r=\sqrt{8 \mu (2 \mu -9)+105}$. This point exists for $\lambda\neq 0, \mu\notin\{1, 2\},$ has eigenvalues denoted symbolically by $\{\delta_1,\delta_2,\delta_3\}$, and is a saddle for
        \begin{enumerate}
             \item $0\leq \mu <1,$ or
             \item $1< \mu <2,$ or
             \item $\mu> 2.$
    
     \end{enumerate}
      \item $Sf_2: \big( -\frac{2}{\lambda (2
     \mu +r-9)}, \frac{\sqrt{-\frac{1}{2} \mu^2
     (4 \mu +r-9)-3 (r+9)}}{3 \lambda \left(\mu ^2-3
     \mu +2\right)}, \frac{1}{6} (-2 \mu
     -r+9)\big).$ This point exists for $\lambda\neq 0, \mu\notin\{1, 2\},$ has eigenvalues denoted symbolically by $\{\lambda_1,\lambda_2,\lambda_3 \}$, and is a saddle for
        \begin{enumerate}
             \item $0\leq \mu <1,$ or
             \item $1< \mu <2,$ or
             \item $\mu> 2.$
         \end{enumerate}
     
   \item $Sf_3: ( 0, 0,
     \frac{1}{6} (-2 \mu -r+9)).$ This point always exists; it has eigenvalues  \newline $\left\{\frac{1}{6} (-2 \mu -r+9),\frac{1}{2} (-\mu(4 \mu +r-21)+3 r-31), \; \frac{1}{6} (-2 \mu (8
     \mu +r-36)+9 r-105)\right\}$. \newline It is: 
     \begin{enumerate}
         \item a source for $1<\mu<2$ and
         \item a sink for $0\leq \mu <1$ or $\mu >2.$
     \end{enumerate}
      \item $Sf_4: \big(
     \frac{2}{\lambda(-2 \mu+r+9)},
     -\frac{\sqrt{\frac{1}{2} \mu ^2 (-4 \mu +r+9)+3
     (r-9)}}{3 \lambda \left(\mu ^2-3 \mu
     +2\right)},\frac{1}{6} (-2 \mu
     +r+9)\big).$ This point exists for $\lambda\neq 0, \mu\notin\{1, 2\},$ has eigenvalues symbolically denoted by $\{\lambda_1,\lambda_2,\lambda_3 \}$,  and is a sink (see Figure \ref{fig:sf5,6}).
      \item $Sf_5: \big(\frac{2}{\lambda (-2
     \mu +r+9)},\frac{\sqrt{\frac{1}{2} \mu^2
     (-4 \mu +r+9)+3 (r-9)}}{3 \lambda \left(\mu ^2-3
     \mu +2\right)}, \frac{1}{6} (-2 \mu
     +r+9)\big).$ This point exists for $\lambda\neq 0, \mu\notin\{1, 2\},$ has eigenvalues denoted symbolically by $\{\overline{\lambda}_1, \overline{\lambda}_2,\overline{\lambda}_3\}$, and is a sink (see Figure \ref{fig:sf5,6}).
      \item $Sf_6: ( 0, 0,
     \frac{1}{6} (-2 \mu +r+9)).$ This point always exists; it has eigenvalues \newline $\left\{\frac{1}{6} (-2 \mu +r+9),\frac{1}{2} (\mu (-4 \mu +r+21)-3 r-31), \; \frac{1}{6} (2 \mu (-8 \mu +r+36)-3 (3
     r+35))\right\}$ and is a saddle for $\mu\geq 0.$
\end{enumerate}

\begin{figure}[ht!]
    \includegraphics[scale=1.0]{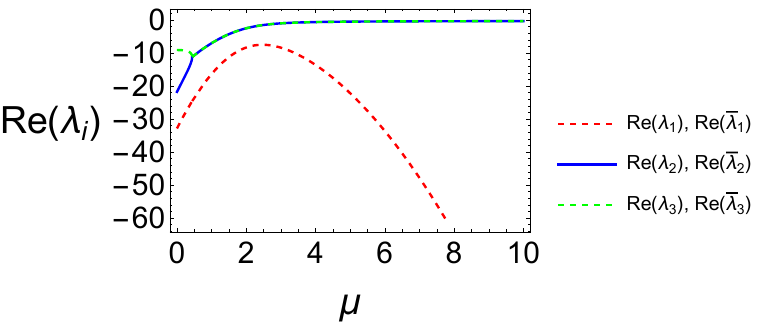}
    \caption{Real part of the eigenvalues $\lambda_1,\lambda_1,\lambda_2$ and $\overline{\lambda}_1,\overline{\lambda}_1,\overline{\lambda}_2$ of $Sf_4$ and $Sf_5$ with $\epsilon=1.$ Both points are sinks.}
    \label{fig:sf5,6}
\end{figure}

In the previous list, the eigenvalues of the points $Sf_1$ and $Sf_4, Sf_5$ are presented symbolically because the expressions are quite complicated. For example, in Figure \ref{fig:sf5,6}, it is verified that both points are sinks. 

\begin{table}[ht!]
 \caption{The best-fit values $(\mu^*, t_U^* )$ for different priors of $\mu$ derived in \cite{Garcia-Aspeitia:2022uxz}.  }
    \label{tab:BEST-FIT}
    \setlength{\tabcolsep}{16.6mm}
\begin{tabular}{ccc}\toprule
 \textbf{Prior}    &  \boldmath{$\mu^*$} & \boldmath{$t_U^*$}\\\midrule
 $0< \mu<1$  & $0.50$ & $41.30\; \text{Gyrs}$ \\
 $ 1 < \mu < 3$ & $ 1.71 $ & $27.89\; \text{Gyrs}$ \\
$ 0 < \mu< 3$ & $1.15$ &$33.66\; \text{Gyrs}$\\\bottomrule
\end{tabular}
\end{table}

On the other hand, the best-fit $\mu$-value is obtained from the reconstruction of $H(z)$ using Equations \eqref{eq:c1_newH}, \eqref{tz}, and \eqref{alternativeE}. The best-fit values $(\mu^*, t_U^* )$ for different priors of $\mu$ were derived in \cite{Garcia-Aspeitia:2022uxz}, and are summarized in Table~\ref{tab:BEST-FIT}. 
These best-fit values are used in the numerics. 
Hence, the general behaviour of the points is verified numerically in Table~\ref{tab:Eigenvalueswithoptimalmu}, where $\mu^*=1.71$ is set for $\epsilon=1$ and $\lambda=4$. 
\begin{table}[ht!]
\caption{Eigenvalues of the Jacobian matrix of the dynamical system \eqref{s1}, \eqref{s2}, \eqref{s3} evaluated at the equilibrium points of Table \ref{tab:critical points} for the best-fit  value $\mu^ *=1.71$, $\epsilon=1$, $\lambda=4$.}
\label{tab:Eigenvalueswithoptimalmu}
\setlength{\tabcolsep}{6.25mm}
\begin{tabular}{cccc}
\toprule
\textbf{Labels} & \boldmath{$\lambda_1$}& \boldmath{$\lambda_2$}& \boldmath{$\lambda_3$}\\\midrule
$Sf_1$ & $0.201645$ & $0.103796$ & $-0.0436487$ \\\midrule
$Sf_2$ & $0.201645$ & $0.103796$ & $-0.0436487$ \\\midrule
$Sf_3$ & $0.0376622$ & $0.0601471$ & $0.201645$ \\\midrule
$Sf_4$ & $-9.75684$ & $-3.42327-3.63794 i$ & $-3.42327+3.63794 i$ \\\midrule
$Sf_5$ & $-9.75684$ & $-3.42327-3.63794 i$ & $-3.42327+3.63794 i$ \\\midrule
$Sf_6$ & $1.82234$ & $-6.84655$ & $-9.75684$
\\ \bottomrule
\end{tabular}%
\end{table}

\begin{figure}[ht!]
   \includegraphics[width=0.45\textwidth]{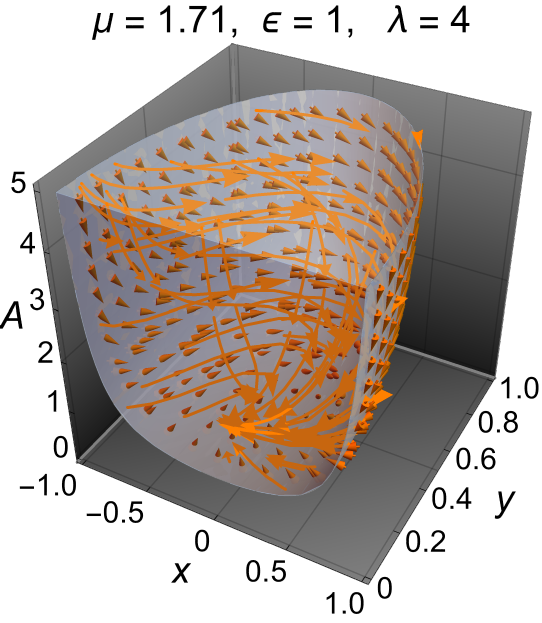}
   \includegraphics[width=0.45\textwidth]{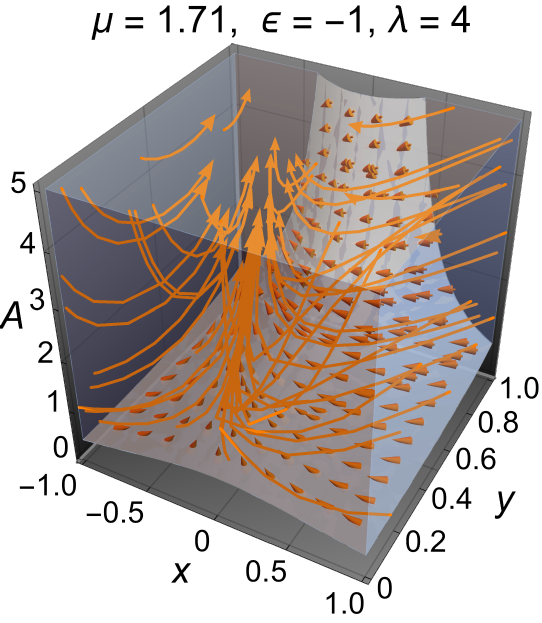}
 \caption{Flow of system  \eqref{s1}, \eqref{s2}, \eqref{s3} for $\epsilon=1$ and $\epsilon=-1$.}
 \label{f:poster}
\end{figure}

 Figure \ref{f:poster} shows the flow of the dynamical system \eqref{s1}, \eqref{s2}, \eqref{s3} for the best-fit value $\mu^*=1.71$.
\subsubsection{Stability Analysis for $\epsilon = -1$}
\begin{figure}[ht!]
    \includegraphics[scale=1.0]{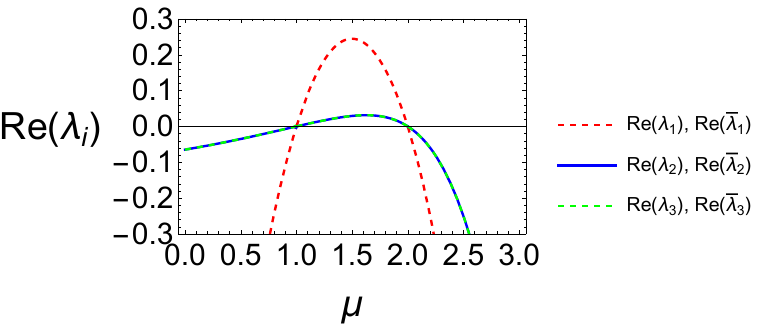}
    \caption{Real part of the eigenvalues of $Sf_1$ and $Sf_2$ for $\epsilon=-1$. Note that $Sf_1$ and $Sf_2$ are sources for $1<\mu<2$ and sinks otherwise. }
    \label{Fig7}
\end{figure}

\begin{figure}[ht!]
    \includegraphics[scale=1.0]{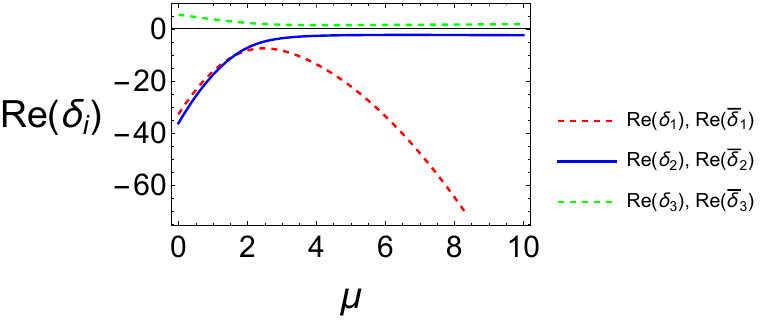}
    \caption{Real part of the eigenvalues of $Sf_4$ and $Sf_5$ for $\epsilon=-1$. Note that $Sf_4$ and $Sf_5$ are saddle points.}
    \label{Fig8}
\end{figure}

The equilibrium points of the system \eqref{s1}, \eqref{s2}, \eqref{s3} for $\epsilon=-1$ with $A\neq 0$ are the following: 
\begin{enumerate}
      \item
      $ Sf_1: \big(-\frac{2}{\lambda (2 \mu +r-9)},-\frac{\sqrt{\frac{1}{2} \mu ^2 (4 \mu + r-9)+3 (r+9)}}{3
     \lambda \left(\mu ^2-3 \mu +2\right)}, \frac{1}{6} (-2 \mu -r+9)\big)$, where \newline $r=\sqrt{8 \mu(2 \mu -9)+105}$. This point exists for $\lambda\neq 0, \mu \notin\{1,2\}$, has eigenvalues symbolically denoted by $\{\lambda_1,\lambda_2,\lambda_3\}$, and (see Figure \ref{Fig7}) is: 
     \begin{enumerate}
         \item a source for $1<\mu<2$
         \item a sink otherwise.
     \end{enumerate}
      \item $Sf_2: \big(-\frac{2}{\lambda (2 \mu
     +r-9)},\frac{\sqrt{\frac{1}{2} \mu ^2 (4 \mu +r-9)+3 (r+9)}}{3 \lambda \left( \mu ^2-3 \mu +2\right)},
     \frac{1}{6} (-2 \mu -r+9)\big).$ This point exists for $\lambda\neq 0, \mu \notin\{1,2\}$, has symbolically denoted eigenvalues times $\{\overline{\lambda}_1,\overline{\lambda}_2,\overline{\lambda}_3\}$, and (see Figure \ref{Fig7}) is: 
     \begin{enumerate}
         \item a source for $1<\mu<2$
         \item a sink otherwise.
     \end{enumerate}
      \item $ Sf_3=(0, 0, \frac{1}{6} (-2 \mu -r+9))$: this point always exists, has eigenvalues \newline $\left\{\frac{1}{ 6} (-2 \mu -r+9),\frac{1}{2} (-\mu
      (4 \mu +r-21)+3 r-31), \; \frac{1}{6} (-2 \mu (8
     \mu +r-36)+9 r-105)\right\}$,\newline and is: \begin{enumerate}
         \item a source for $1<\mu<2$
         \item a sink for $0\leq \mu <1$ or $\mu >2.$
     \end{enumerate}
   \item $Sf_4: \big(
     \frac{2}{\lambda (-2 \mu +r+9)}, -\frac{\sqrt{2 \mu^3-\frac{1}{2} \mu^2 (r+9) -3 (r-9)}}{3 \lambda \left(\mu
     ^2-3 \mu +2\right)}, \frac{1}{6} (-2 \mu +r+9)\big).$ This point exists for $\lambda\neq 0, \mu \notin\{1,2\}$, has eigenvalues denoted symbolically by $\{\delta_1,\delta_2,\delta_3\}$, and is a saddle (see Figure \ref{Fig8}).
      \item $ Sf_5: \big(\frac{2}{\lambda (-2 \mu +r+9)},
     \frac{\sqrt{2 \mu ^3-\frac{1}{2} \mu ^2 (r+9)-3 (r-9)}}{3 \lambda \left(\mu ^2- 3 \mu  +2\right)}, \frac{1}{6}
     (-2 \mu +r+9)\big).$ This point exists for $\lambda\neq 0, \mu \notin\{1,2\}$, has eigenvalues denoted symbolically by $\{\overline{\delta}_1,\overline{\delta}_2,\overline{\delta}_3\}$, and is a saddle (see Figure \ref{Fig8}).
     \item $Sf_6: (0, 0,\frac{1}{6} (-2 \mu +r+9)).$ This point always exists, has eigenvalues \newline $\left\{\frac{1} {6} (-2 \mu  +r+9),\frac{1}{2} (\mu 
     (-4 \mu  +r+21)-3 r-31), \; \frac{1}{6} (2 \mu  (-8
     \mu +r+36)-3 (3 r+35))\right\}$, and is a saddle.
\end{enumerate}

As in the previous section, the eigenvalues of the points $Sf_{1}$, $Sf_{2}$, $Sf_{4}$, and $Sf_{5}$ have been written symbolically; however, we studied stability of these points numerically as well. In Figure~\ref{Fig7}, it is verified that $Sf_{1}$ and $Sf_{2}$ are sources or sinks, while in Figure  \ref{Fig8} it is illustrated that $Sf_{4}$ and $Sf_{5}$ are saddles. On the other hand, the general behaviour of the points is verified numerically in Table \ref{tab:Eigenvalueswithoptimalmu2}, where we set $\mu^*=1.71$, $\epsilon=-1$, and $\lambda=4$.

\begin{table}[ht!]
\caption{Eigenvalues of the Jacobian matrix of the dynamical system \eqref{s1}, \eqref{s2}, \eqref{s3} evaluated at the equilibrium points of Table \ref{tab:critical points} for the best-fit value $\mu^ *=1.71$, $\epsilon=-1$, $\lambda=4$.}
\label{tab:Eigenvalueswithoptimalmu2}
\setlength{\tabcolsep}{5.3mm}
\begin{tabular}{cccc}
\toprule
\textbf{Labels} & \boldmath{$\lambda_1$}& \boldmath{$\lambda_2$}& \boldmath{$\lambda_3$}\\\midrule
$Sf_1$ & $0.201645$  & $0.0300736 +0.0602174 i$ & $0.0300736
   -0.0602174 i$  \\\midrule
$Sf_2$ &  $0.201645$ 
 & $0.0300736 +0.0602174 i$ & $0.0300736
   -0.0602174 i$  \\\midrule
$Sf_3$ & $0.0376622$ & $0.0601471$ & $0.201645$  \\\midrule
$Sf_4$ & $-9.75684$ & $-9.47903$ & $2.63249$ \\\midrule
$Sf_5$ & $-9.75684$ & $-9.47903$ & $2.63249$  \\\midrule
$Sf_6$ & $1.82234$ & $-6.84655$ & $-9.75684$
\\ \bottomrule
\end{tabular}%

\end{table}

\subsection{Bianchi Metric I}

The local and rotationally symmetric Bianchi I spacetime can be described by the line element \eqref{ch.03}. Then, the effective fractional action \eqref{LFCBI} can be generalized by incorporating a scalar field and a perfect fluid with energy density $\rho$ and pressure $p$: 
\begin{align}
&  S_{\text{eff}}  = \frac{1}{\Gamma(\mu)}\int_0^t \Bigg[\frac{3}{8\pi G N^2(\theta)}\Bigg(\ddot{\alpha}(\theta)-\frac{\dot{\alpha}(\theta)\dot{N}(\theta)}{N(\theta)}%
+2\dot{\alpha}^{2}(\theta)+\frac{1}{4}\dot{\beta}^{2}(\theta)\Bigg) \nonumber\\ 
    &\;\;\;\;\;\;\;\;\;\;\;\;\;\;\;\;\;\;\;\;\;\;\;\;\;\; +e^{3\alpha(\theta)}\mathcal{L}_{\text{m}} +e^{3\alpha(\theta)}\left(\frac{\epsilon \Dot{\phi}^2(\theta)}{2N^2(\theta)}-V(\phi(\theta))\right)\Bigg]  (t-\theta)^{\mu-1} N(\theta) d\theta\label{LFCBI-sf}. 
\end{align}

Proceeding with the variation of the action, we define $H=\dot{\alpha}$ and $\sigma=\dot{\beta}/2$, the Hubble parameter and the anisotropy parameter, respectively. 
Then, the following system is obtained:
\begin{align}
&    \Ddot{\phi}+3\left(H+\frac{1-\mu}{3t}\right)\Dot{\phi}+\epsilon\frac{dV(\phi)}{d\phi}=0,\\
 &   \Dot{H}-\frac{1-\mu}{2t}H+\frac{(1-\mu)(2-\mu)}{2t^2} + 3 \sigma^2 =-4\pi G \left(\epsilon \Dot{\phi}^2+\rho_m + p_m\right), \\
& \dot{\sigma}+ 3\sigma \left(H+\frac{1-\mu }{3 t}\right)=0, \\
& H^2 +\frac{(1-\mu) H}{t} -\sigma^2=\frac{8\pi G}{3}\left(\epsilon \frac{\Dot{\phi}^2}{2}+V(\phi)+\rho_m \right)+\frac{\Lambda}{3},
\end{align}
along with the continuity equation
\begin{equation}
    \Dot{\rho}_m+3\left(H+\frac{1-\mu}{3t}\right)(\rho_m +p_m)=0.
\end{equation}

To designate the temporary independent variables, the rule $t-\theta \mapsto t$, $\theta \mapsto t$ \cite{Shchigolev:2010vh} is used, where the dots denote these derivatives. 
Again, dimensionless variables \eqref{sf-vars} are defined together with \eqref{AlternativevarsBI},
which satisfies the following condition defining $\Omega_m $: 
\begin{equation}
   1-A \left(x^2 \epsilon
   +y^2 -1\right)-\mu
   :=A \left(\Sigma ^2+\Omega_m + \Omega_\Lambda \right)\geq 0.
   \label{condición variables Bianchi I}
\end{equation}

Using the derivative $f^{\prime}=A^2 \dot{f}/H$, we obtain
\begin{align}
x^{\prime}&= 3 A^2 \lambda  y^2 \epsilon +3 A (\mu -3) x-(\mu -2) (\mu-1) x,     \label{xprime}  \\
y^{\prime}&= -y  \left(-2 A \mu +A (3 A (\lambda  x-1)+8)+\mu^2-3 \mu +2\right),     \label{yprime}\\
A^{\prime}&=A \left(-2 A \mu -3 (A-3) A+\mu ^2-3 \mu+2\right),  \label{Aprime} 
\\
\Omega_{\Lambda}^{\prime}&=-2 \Omega_\Lambda  (A (-3 A-2 \mu+8)+(\mu -2) (\mu -1)), \label{OmegaLambdaprime}
\\
\Sigma^{\prime}&= \Sigma  (3 A (\mu -3)-(\mu-2) (\mu -1)). \label{Sigmaprime}
\end{align}

The equilibrium points of the system \eqref{xprime}, \eqref{yprime}, \eqref{Aprime}, \eqref{OmegaLambdaprime}, \eqref{Sigmaprime} are shown in Table \ref{tab:critical points X}. It can be observed that the first three equations and the last three equations are decoupled. Thus, two uncoupled subsystems can be studied: the state vector $(x,y,A)$, which evolves according to  \eqref{xprime},  \eqref{yprime}, \eqref{Aprime}, and on the other hand the state vector $(A, \Omega_\Lambda, \Sigma)$, which evolves according to \eqref{Aprime}, \eqref{OmegaLambdaprime}, and \eqref{Sigmaprime}.
As in the previous sections, due to physical considerations we do not examine the points with $A=0$. Moreover, we assume $\mu\notin\{1,2\}$ for the parameter space. 
\begin{table}[ht!]
\caption{Equilibrium points of the dynamical system \eqref{xprime}, \eqref{yprime}, \eqref{Aprime}, \eqref{OmegaLambdaprime}, \eqref{Sigmaprime}, where $r=\sqrt{8\mu(2\mu -9)+105}$.}
\label{tab:critical points X}
\setlength{\tabcolsep}{3.5mm}
\begin{tabular}{cccccc}
\toprule
\textbf{Labels} & \boldmath{$x$} & \boldmath{$y$} & \boldmath{$A$} & \boldmath{$\Omega_\Lambda$}  & \boldmath{$\Sigma$}\\
\midrule
$Sf_1$ &  $-\frac{2}{\lambda  \left(2 \mu +r-9\right)}$ & $-\frac{\sqrt{-\left(\left(4 \mu +r-9\right) \mu ^2\right)-6\left(r+9\right)}}{3\sqrt{2} \sqrt{\lambda ^2 \left(\mu ^2-3 \mu+2\right)^2 \epsilon }}$ & $\frac{1}{6}\left(-2 \mu -r+9\right)$    & $0$   & $0$ \\ [0.9ex]
\midrule
$Sf_2$ & $-\frac{2}{\lambda  \left(2\mu +r-9\right)}$ & $\frac{\sqrt{-\left(\left(4 \mu +r-9\right) \mu ^2\right)-6 \left(r+9\right)}}{3 \sqrt{2} \sqrt{\lambda ^2 \left(\mu ^2-3 \mu+2\right)^2 \epsilon }}$ & $\frac{1}{6} \left(-2 \mu-r+9\right)$ & $0$  & $0$ \\ [0.9ex]
\midrule
$Sf_3$ & $0$ & $0$ & $\frac{1}{6} \left(-2 \mu -r+9\right)$ & $0$  & $0$ \\ [0.9ex]
\midrule
$Sf_4$ & $\frac{2}{\lambda  \left(-2 \mu +r+9\right)}$  & $-\frac{\sqrt{\left(-4 \mu +r+9\right) \mu ^2+6 \left(r-9\right)}}{3 \sqrt{2} \sqrt{\lambda^2 \left(\mu ^2-3 \mu +2\right)^2 \epsilon}}$ & $\frac{1}{6} \left(-2 \mu+r+9\right)$    & $0$  & $0$ \\ [0.9ex]
\midrule
$Sf_5$ & $\frac{2}{\lambda  \left(-2 \mu +r+9\right)}$  & $\frac{\sqrt{\left(-4\mu +r+9\right) \mu^2+6 \left(r-9\right)}}{3 \sqrt{2} \sqrt{\lambda ^2\left(\mu ^2-3 \mu +2\right)^2 \epsilon}}$ & $\frac{1}{6}\left(-2 \mu +r+9\right)$    & $0$  & $0$ \\ [0.9ex]
\midrule
$Sf_6$ & $0$  & $0$  & $\frac{1}{6}\left(-2 \mu +r+9\right)$    & $0$ & $0$ \\ [0.9ex]
\bottomrule
\end{tabular}%

\end{table}

\subsubsection{First Uncoupled System: Dynamics in Subspace $\left(x, y, A \right)$}
In this case, the system turns out to be precisely the same as the system \eqref{s1}, \eqref{s2}, \eqref{s3}; see the analysis in Section \ref{Sf_First.System}.

\subsubsection{Second Decoupled System: Dynamics in Subspace $\left(A\right.,\Omega_{\Lambda}$,$\left.\Sigma\right)$}

The equilibrium points of the decoupled system
provided by the equations \eqref{Aprime}, \eqref{OmegaLambdaprime}, and \eqref{Sigmaprime} that satisfy $A\neq 0$ are:
\begin{enumerate}
      \item $T_1: (\frac{1}{6} \left(-2 \mu -r+9\right),0,0).$ This point always exists and is
      \begin{enumerate}
          \item a source for $1<\mu < 2$
          \item a sink for $0\leq \mu <1$ or $\mu > 2$.
      \end{enumerate}
      \item $T_2: (\frac{1}{6} \left(-2 \mu +r+9\right),0,0).$ This point always exists and is a saddle for $\mu \geq 0$.
\end{enumerate}

Table \ref{tab:Criticalpointsofuncoupledsystem1} shows the equilibrium points of the uncoupled system \eqref{Aprime}, \eqref{OmegaLambdaprime}, \eqref{Sigmaprime} with $A\neq 0$, where $r=\sqrt{8\mu(2\mu-9)+105}$. 

\begin{table}[ht!]
\caption{Equilibrium points of the decoupled system \eqref{Aprime}, \eqref{OmegaLambdaprime}, \eqref{Sigmaprime}, where $r=\sqrt{8\mu(2\mu-9)+105}$.}
\label{tab:Criticalpointsofuncoupledsystem1}
\setlength{\tabcolsep}{5.85mm}
\begin{tabular}{ccccc}
\toprule
\textbf{Labels} & \boldmath{$A$}& \boldmath{$\Omega_{\Lambda}$}& \boldmath{$\Sigma$} & \textbf{Stability}\\\midrule
$T_1$ & $\frac{1}{6} \left(-2 \mu -r+9\right)$ & $0$ & $0$ & Sink or source (see text).\\ \midrule
$T_2$ & $\frac{1}{6} \left(-2 \mu +r+9\right)$ &  $0$ &    $0$ & Saddle\\\bottomrule
\end{tabular}

\end{table}

Table \ref{tab:Eigenvaluesdecoupledsystem1} shows the eigenvalues of the Jacobian matrix of the dynamical system \eqref{Aprime}, \eqref{OmegaLambdaprime}, \eqref{Sigmaprime} evaluated at the equilibrium points in Table \ref{tab:Criticalpointsofuncoupledsystem1}.

\begin{table}[ht!]
\caption{Eigenvalues of the Jacobian matrix of the dynamical system \eqref{Aprime}, \eqref{OmegaLambdaprime}, \eqref{Sigmaprime} evaluated at the equilibrium points in Table \ref{tab:Criticalpointsofuncoupledsystem1}, where $r=\sqrt{8 \mu(2\mu-9)+105}$.}
\label{tab:Eigenvaluesdecoupledsystem1}
\resizebox{1\textwidth}{!}{%
\begin{tabular}{cccc}
\toprule
\textbf{Labels} & \boldmath{$\lambda_1$} & \boldmath{$\lambda_1$} & \boldmath{$\lambda_3$}\\ \midrule
   $T_1$ &$\frac{1}{3} (-2 \mu -r+9)$ & $\frac{1}{2} (-\mu  (4 \mu +r-21)+3 r-31)$& $\frac{1}{6} (-2 \mu  (8 \mu +r-36)+9 r-105)$ \\\midrule
   $T_2$ &  $\frac{1}{3} (-2 \mu +r+9)$ & $\frac{1}{2} (\mu  (-4 \mu +r+21)-3 r-31)$ & $\frac{1}{6} (2 \mu  (-8 \mu +r+36)-3 (3
   r+35))$ \\\bottomrule
\end{tabular}%
}

\end{table}

\begin{figure}[ht!]
    
    \includegraphics[scale=0.8]{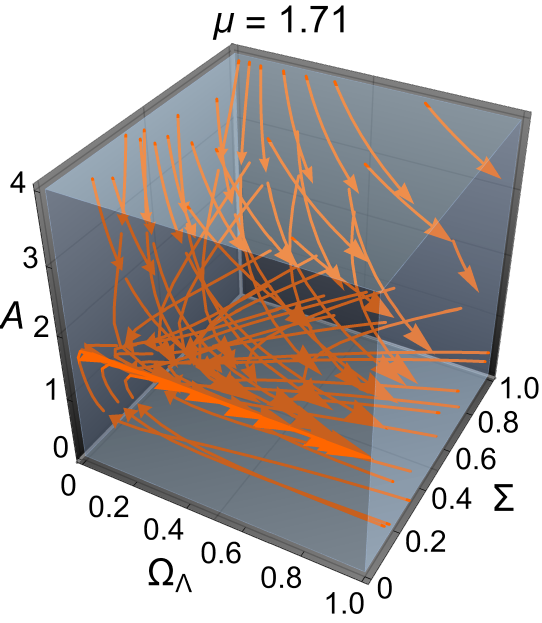}
    \caption{Flow of the dynamical system \eqref{Aprime}, \eqref{OmegaLambdaprime}, \eqref{Sigmaprime} for  $\mu^*=1.71$.}
    \label{fig:9}
\end{figure}

Figure \ref{fig:9} shows the flow of the dynamical system \eqref{Aprime}, \eqref{OmegaLambdaprime}, \eqref{Sigmaprime} for  $\mu^*=1.71$.
\section{Results}
\label{RRResults}

In \cite{Garcia-Aspeitia:2022uxz}, the recent proposal of fractional cosmology was studied and the theory was found to correctly predict the acceleration of the universe, thereby providing clues about the fundamental nature of dark energy. By writing Einstein's field equations in the fractional formulation, the Friedmann equations naturally contain a constant term predicting the existence of an accelerating late universe with only matter and radiation. This is contrary to the standard approach, in which it is necessary introduce a cosmological constant \cite{Garcia-Aspeitia:2022uxz}. 

To complement the results achieved in \cite{Garcia-Aspeitia:2022uxz}, in the present investigation Section \ref{alternattive_I} is dedicated
to discussing an alternative study using the Riccati Equation \eqref{NewIntegrableH} and assuming that the matter components have the equation of state $p_i=w_i\rho_i$, where $w_i \neq -1$ are constants. For example, for the flat FLRW metric it is observed that Equations \eqref{FriedmannFrac} and \eqref{Newpressure} impose restrictions on the type of matter components of the universe, say, 
\begin{align}
    H^2+\frac{(1-\mu)}{t}H &=\frac{8\pi G}{3}\sum_i\rho_i, \label{ALIg11}
\\
  \frac{2 (\mu -3) H}{t}+H^2-\frac{ (\mu -2) (\mu -1)}{t^2} &= \frac{8 \pi G}{3} \sum_i w_i \rho_i. \label{ALIg22}
\end{align}

The analysis of the second constraint was not developed in \cite{Garcia-Aspeitia:2022uxz}. This constraint is an immediate consequence of the Riccati Equation \eqref{NewIntegrableH} and of considering the accepted cosmological hypothesis that the conservation equations of the different matter components are separately conserved.
Constraints \eqref{ALIg11} and \eqref{ALIg22} are written in dimensionless form as
\begin{align}
 1+\frac{(1-\mu)}{A} &=\sum_i\Omega_i, \label{LIg11}
\\
1+ \frac{2 (\mu -3)}{A}-\frac{ (\mu -2) (\mu -1)}{A^2}& = \sum_i w_i \Omega_i,\label{LIg22}
\end{align}
where $A:=t H$ is the known age parameter and $\Omega_i$ represents the dimensionless densities of the different matter components of the universe. 

At this point, there are two routes that can be taken to further investigate the model without a scalar field: (i) it is imposed that the components of the Universe are CDM ($\rho_{1}=\rho_ {\text{m}}, w_1=w_{\text{m} }=0$) and radiation ($\rho_{2}=\rho_{\text{r}}, w_2=w_{\text{r }}=1/3$); alternatively, (ii) the equation of state of one of the matter sources is not imposed, and is rather deduced from the compatibility conditions \eqref{LIg11} and \eqref{LIg22}.

In the first case, from the constraints \eqref{LIg11} and \eqref{LIg22} and the conditions of existence of the equilibrium points/sets of the system \eqref{EQ3.45}, \eqref{EQ3.46}, \eqref{EQ3.47} that satisfy the compatibility conditions \eqref{LigB} and \eqref{LigNewB} (which are deduced from the constraints \eqref{LIg11} and \eqref{LIg22}), we can obtain the possible values of $\mu$. To be more precise, in Section \ref{alternattive_I} the following novel result is obtained: considering that the matter sources are radiation and cold dark matter and that $\mu\notin\{1,2\}$, this imposes conditions on the parameter $\mu$, which can take only the discrete values $\{5/2, 7/2, 41/8\}$. The system is then reduced to a one-dimensional system provided by \eqref{EQ3.47} for these values.

Table \ref{tab:P} shows the equilibrium points/sets of the system \eqref{EQ3.45}, \eqref{EQ3.46}, \eqref{EQ3.47} that satisfy the compatibility conditions \eqref{LigB} and \eqref{LigNewB}. We have omitted the analysis of the points with $A=0$ because they are not physically interesting. Recall that for equilibrium points with constant $A$ the corresponding cosmological solution is a power-law solution with scale factor $a(t)= \left(t/t_U\right)^{A}$. Then, the solutions $P_1$ and $P_2$ verify that $a(t)= \left(t/t_U\right)^{\frac{3}{2}}$. Finally, the solution $P_3$ satisfies $a(t)= \left(t/t_U\right)^\frac{15}{8}$. However, points $P_2$ and $P_3$ are nonphysical, as they lead to $\Omega_{\text{m}}<0$ and $\Omega_{\text{r}}<0$, respectively.

Following the second alternative route, the components of the universe are assumed to be CDM ($\rho_{1}=\rho_{\text{m}}, w_1=w_{\text{m} }=0$) and a fluid with its constant equation of state to be determined ($\rho_{2}=\rho_{\text{X}}, w_2=w_{\text{X}}$). Because we have a free parameter $w_{\text{X}}$, we can obtain the values from the unspecified fluid equation of state, which provides the acceleration of the expansion without considering the cosmological constant or a scalar field. Table \ref{tab:R2} presents the equilibrium points/sets of the system \eqref{EQ3.61}, \eqref{EQ3.62}, \eqref{EQ3.63} that satisfy the compatibility conditions \eqref{Lig2B} and \eqref{LigNew2B} deduced from the constraints \eqref{LIg11} and \eqref{LIg22}. Hence, we obtain the solution $R_5: \Big\{A= \frac{1}{6} \left(-2 \mu +r+9\right),\Omega_{\text{m}}= 0, \Omega_{\text{X }}= -\frac{r-5}{2 (\mu -2)}\Big\}$, where $ r=\sqrt{8 \mu (2 \mu -9)+105}$. The equation of state of the effective fluid ($w_{\text{X}}$) and the deceleration parameter ($q$) of the equilibrium point $R_5$ are $w_{\text{X}}= \frac{r-7}{4 (\mu -1)}$  and  $q= \frac{-2 (\mu -4) \mu +r-13}{2 (\mu -2) (\mu -1 )}$, respectively. Figure \ref{fig:R_13} shows $w_{\text{X}}$  and $q$ for equilibrium point $R_5$ as a function of $\mu$ in the physical region within the parameter space, and $w_{\text{X}}$ satisfies $-1<w_{\text{X}}<-1/3$. The extra fluid mimics the equation of state of a quintessence scalar field. Now, we have an attractor solution that accelerates the expansion without considering the cosmological constant or a scalar field.
The equilibrium point is:
\begin{itemize}
\item nonhyperbolic, with a two-dimensional stable manifold for $\mu <\frac{7}{2}$
\item a saddle for $\mu >\frac{7}{2}$
\item has a solution that satisfies $w_{\text{X}}<-1/3$ for $\mu <\frac{5}{2}$
\item the solution is accelerated ($q<0$) for large $t$ for all $\mu$
\item The solution is not physically viable for $\mu >\frac{5}{2}$.
\end{itemize}

Another line of research is to consider a spacetime that is both homogeneous and anisotropic, particularly the Bianchi I metric. The compatibility conditions are obtained using the dimensionless variables \eqref{AlternativevarsFLRW} and \eqref{AlternativevarsBI}, say, 
\begin{align}
   1+\frac{(1-\mu)}{A} -\Sigma^2 &=\sum_i\Omega_i, \label{NNNB1}
\\
1+ \frac{2 (\mu -3)}{A} -\frac{ (\mu -2) (\mu -1)}{A^2} -\Sigma^2 &= \sum_i w_i \Omega_i , \label{NNNB2}
\end{align}
where $\Sigma$ is a dimensionless measure of the spacetime's anisotropy. When $\Sigma\rightarrow 0$, the compatibility conditions \eqref{LIg11} and \eqref{LIg22} are retrieved.

As before, there are two routes in the investigation for the model without a scalar field: (i) it is imposed that the components of the Universe are  CDM and radiation; alternatively, (ii) the equation of state of one of the matter sources is not imposed, and is instead deduced from the compatibility conditions.

Then, considering that the matter sources are radiation and cold dark matter and that $\mu\notin\{1,2\}$, this imposes conditions on the parameter $\mu$, which can take only the discrete values $\{5/2, 7/2, 41/8\}$. In the first case, the equilibrium points/sets of the system \eqref{EQ3.86}, \eqref{EQ3.87}, \eqref{EQ3.88}, \eqref{EQ3.89} that satisfy the compatibility conditions \eqref{LigC} and \eqref{LigNewC} (which are derived from \eqref{NNNB1} and \eqref{NNNB2}) and $A\neq0$ are presented in Table \ref{tab:S}.

Following the second alternative route, Table \ref{tab:S2} presents the equilibrium points/sets of the system \eqref{EQ3.113}, \eqref{EQ3.114}, \eqref{EQ3.115}, \eqref{EQ3.116} that satisfy the compatibility conditions \eqref{Lig3B} and \eqref{LigNew3B} that are deduced from \eqref{NNNB1} and \eqref{NNNB2}. As before, there exists a cosmological solution in the physical region within the parameter space $\mu \in \left(-\infty, \frac{5}{2}\right]$ and for which equation of state  $w_{\text{X}}$ satisfies $-1<w_{\text{X}}<-1/3$, that is, the extra fluid mimics the equation of state for quintessence. As such, we have an attractor solution that accelerates the expansion without considering the cosmological constant or a scalar field.

The solution $R_5: \left\{\Omega_{\text{m}}= 0,\Omega_{\text{X}}= -\frac{r-5}{2 (\mu -2)}, \Sigma = 0,A= \frac{1}{6}
   \left(-2 \mu +r+9\right)\right\}$ has the cosmological parameters  $q= \frac{-2 (\mu -4) \mu +r-13}{2 (\mu -2) (\mu -1)}$ and $w_{\text{X} }= \frac{r-7}{4 (\mu -1)}$.
The equilibrium point is:
\begin{itemize}
\item nonhyperbolic with a three-dimensional stable manifold for $\mu <\frac{7}{2}$
\item is a saddle for $\mu >\frac{7}{2}$
\item has a solution that satisfies $w_{\text{X}}<-1/3$ for $\mu  <\frac{5}{2}$
\item the solution is accelerated ($q<0$) for large $t$ and all $\mu$
\item the solution is not physically viable for $\mu >\frac{5}{2}$.
\end{itemize}

Moreover, a more general model can be shown by incorporating a scalar field and the cosmological constant $\Lambda$ as matter sources. The scalar field $\phi$ has an exponential potential $V(\phi)=V_0 e^{-4 \sqrt{3 \pi } \sqrt{G} \lambda \phi }$. The kinetic energy is $ \epsilon {\dot{\phi}}^2/2$, where $\epsilon=\pm 1$ depending on whether the scalar field has positive (quintessence) or negative ({phantom}) kinetic energy. 

This article discusses the physical interpretation of the corresponding cosmological solutions, with particular emphasis on the influence of the order of the fractional derivative on the theory. Our results improve and extend previous results reported in the literature. 

The quintessence model in the flat FLRW metric is described by the system \eqref{s1}, \eqref{s2}, \eqref{s3} for $\epsilon=1$. The past and future attractors for this model are the following:
\begin{enumerate}
    \item[$Sf_3$:] always exists and is
   \begin{enumerate}
       \item a source for $1<\mu<2$
       \item a sink for $0\leq \mu <1$ or $\mu >2$.
   \end{enumerate}
\item[$Sf_4$:] exists for $\lambda\neq 0, \mu\notin\{1, 2\}$ and is a sink (see Figure \ref{fig:sf5,6}).

\item [$Sf_5$:] exists for $\lambda\neq 0, \mu\notin\{1, 2\}$ and is a sink (see Figure \ref{fig:sf5,6}).
\end{enumerate}

The {phantom} model in the flat FLRW metric is described by the system \eqref{s1}, \eqref{s2}, \eqref{s3} for $\epsilon=-1$. The past and future attractors for this model are the following:
\begin{enumerate}
    \item[$ Sf_1$:] exists for $\lambda\neq 0, \mu \notin\{1,2\}$ and (see Figure \ref{Fig7}) is:
   \begin{enumerate}
       \item a source for $1<\mu<2$
       \item a sink otherwise.
   \end{enumerate}
   \item[$ Sf_2$:] exists for $\lambda\neq 0, \mu \notin\{1,2\}$ and (see Figure \ref{Fig7}) is:
   \begin{enumerate}
       \item a source for $1<\mu<2$
       \item a sink otherwise.
   \end{enumerate}
     \item[$ Sf_3$:] always exists and is \begin{enumerate}
       \item a source for $1<\mu<2$
       \item a sink for $0\leq \mu <1$ or $\mu >2.$
   \end{enumerate}
\end{enumerate}

Finally, we investigated the model with a scalar field with positive or negative  kinetic energy in the Bianchi I metric, as provided by the system \eqref{xprime}, \eqref{yprime}, \eqref{Aprime}, \eqref{OmegaLambdaprime}, \eqref{Sigmaprime}. The equilibrium points of this system are shown in Table \ref{tab:critical points X}. An important aspect of this model is that the first three equations and the last three equations are decoupled. Thus, two uncoupled subsystems can be studied: the state vector $(x,y,A)$, which evolves according to \eqref{Aprime}, \eqref{xprime}, and \eqref{yprime}, and the state vector $(A, \Omega_\Lambda, \Sigma)$, which evolves according to \eqref{Aprime}, \eqref{OmegaLambdaprime}, and \eqref{Sigmaprime}. These model the system's dynamics in different invariant sets for the flow. In the first case, the system turns out to be the same as the system \eqref{s1}, \eqref{s2}, \eqref{s3}, and the previous results are reproduced (see Section \ref{Sf_First.System}). On the other hand, the equilibrium point of interest of the system \eqref{Aprime}, \eqref{OmegaLambdaprime}, \eqref{Sigmaprime} is $T_1: \left(A,\Omega_{\Lambda }, \Sigma\right)= (\frac{1}{6} \left(-2 \mu -r+9\right),0,0)$. This point always exists, and is
a source for $1<\mu < 2$ and a sink for $0\leq \mu <1$ or $\mu > 2$. It can be confirmed that the solutions isotropize ($\Sigma\rightarrow 0$) at late times.

\section{Conclusions}
\label{conclusiones}

Fractional calculus is a generalization of classical integer order calculus in which derivatives and integrals are of arbitrary order $\mu$. This formalism is used to investigate objects and systems characterized by nonlocality, long-term memory, or fractal properties and derivatives of non-integer orders. In many cases, this approach can model real-world phenomena in a better way than using classical calculus. For example, extensions of the $\Lambda$CDM concordance model can be obtained by modifying General Relativity and introducing fractional cosmology. In this theory, the Friedmann equation is modified and the late-time cosmic acceleration is obtained without incorporating dark energy. The modified theory has been compared in the literature against data from cosmic chronometers, observations of type Ia supernovae, and their joint analysis, allowing the range of $\mu$ \cite{Garcia-Aspeitia:2022uxz} to be restricted.

On the other hand, we carried out an analysis of dynamical systems in order to determine the model's asymptotic states and discover the influence of the parameter $\mu$ on the dynamics. This analysis allows for good understanding of the global structure of the reduced phase spaces. Finally, we explored the phase space for different values of the fractional order of the derivative as well as for different matter models. The objective of the investigation was to classify equilibrium points and provide a range for the fractional order of the derivative in order to obtain a late-term accelerating power-law solution for the scale factor.
With these elements as a starting point, in this paper we have continued the previous work of \cite{Garcia-Aspeitia:2022uxz}. In this sense, two research paths were identified for the model without a scalar field: (i) to compare it with the standard model, it is imposed that the two components of the universe are CDM and radiation; alternatively, (ii) the equation of state of one of the matter sources is not imposed, and is instead deduced from the compatibility conditions. The analysis of this second constraint was not performed in the previous work of \cite{Garcia-Aspeitia:2022uxz}. 
This constraint is an immediate consequence of Riccati's Equation \eqref{NewIntegrableH} and of considering the accepted cosmological hypothesis that the conservation equations of the different matter components are separately conserved.

By incorporating a scalar field as a matter source, these results and previous results from the literature are complemented and generalized by our analysis. The most relevant novel results are discussed in Section \ref{RRResults}. Our results improve upon and extend the previous results in the literature. Consequently, we can affirm that fractional calculus is able to play a relevant role in describing physical phenomena, particularly with respect to theories of gravity. In this approach, traditional (non-fractional) General Relativity can only approximate the mathematical structure that describes nature. It is worth noting the importance of using advanced mathematical methods in theoretical cosmology, which provides fertile ground for new formulations and more prominent tools to reach a better and more meaningful understanding of the universe.

\vspace{6pt} 



\section*{Author Contributions}
{Conceptualization, B.M.-R., A.D.M. and G.L.; methodology, G.L.; software,  B.M.-R., A.D.M. and G.L.; validation,  B.M.-R., A.D.M. and G.L.; formal analysis, B.M.-R., A.D.M., G.L., C.E. and A.P.; investigation, B.M.-R., A.D.M., G.L., C.E. and A.P.; resources, G.L., C.E. and A.P.;  writing---original draft preparation, G.L.; writing---review and editing, G.L., C.E. and A.P.; visualization,  B.M.-R., A.D.M. and G.L.; supervision, G.L. and C.E.; project administration, G.L. and C.E.; funding acquisition, A.D.M., G.L., C.E. and A.P. All authors have read and agreed to the published version of the manuscript.}

\section*{Funding}

C.E. and B.M.-R. were funded by Agencia Nacional de Investigación y Desarrollo (ANID) through Proyecto Fondecyt Iniciación folio 11221063, Etapa 2022. A.D.M. was supported by ANID Subdirección de Capital Humano/Doctorado Nacional/año 2020 folio 21200837, Gastos operacionales proyecto de tesis/2022 folio 242220121, and Vicerrectoría de Investigación y Desarrollo Tecnológico (VRIDT) at Universidad Católica del Norte. G.L. was funded by VRIDT-UCN through Concurso De Pasantías De Investigación Año 2022, Resolución VRIDT No. 040/2022 and Resolución VRIDT No. 054/2022.  A.P. acknowledges the funding of VRIDT-UCN through Concurso de Estadías de Investigación, Resolución VRIDT N°098/2022.

\section*{Data availability}

No new data were created or analyzed in this study. Data sharing is not applicable to this article.

\acknowledgments{The authors are thankful for the support of Núcleo de Investigación Geometría Diferencial y Aplicaciones, Resolución VRIDT No. 096/2022. The authors sincerely thanks three anonymous referees for their bold and encouraging comments. }

\section*{Conflicts of interest}

The authors declare no conflict of interest. The funders had no role in the design of the study; in the collection, analyses, or interpretation of data; in the writing of the manuscript; or in the decision to publish the~results.

\end{document}